\documentclass[a4]{article}

\usepackage[a4paper,top=3cm,bottom=2cm,left=2cm,right=2cm,marginparwidth=1.75cm]{geometry}

\usepackage[english]{babel}
\usepackage[utf8x]{inputenc}
\usepackage[T1]{fontenc}

\usepackage{amsmath}
\usepackage[colorinlistoftodos]{todonotes}
\usepackage[colorlinks=true, allcolors=blue]{hyperref}
\usepackage{url}
\usepackage{hyperref}
\usepackage{multicol}
\usepackage{graphicx}
\usepackage{subfig}

\usepackage{multirow}
\DeclareMathAlphabet {\mathbfit}{OML}{cmm}{b}{it}
\newcommand{\beginsupplement}{%
        \setcounter{table}{0}
        \renewcommand{\thetable}{S\arabic{table}}%
        \setcounter{figure}{0}
        \renewcommand{\thefigure}{S\arabic{figure}}%
     }

\begin{document}

\vspace*{0.35in}

\begin{flushleft}
{\Large
\textbf\newline{Uncovering Multi-Site Identifiability Based on Resting-State Functional Connectomes}
}
\newline
\\
Sumra Bari$^{\textbf{1}}$,
Enrico Amico$^{\textbf{2,3}}$,
Nicole Vike$^{\textbf{4}}$,
Thomas M. Talavage$^{\textbf{1,5}}$,
Joaqu\'{i}n~Go\~{n}i$^{\textbf{2,3,5,*}}$\\
\bigskip
$^{1}$School of Electrical and Computer Engineering, Purdue University, West-Lafayette, IN, USA.\\ 
$^{2}$School of Industrial Engineering, Purdue University, West-Lafayette, IN, USA.\\ 
$^{3}$Purdue Institute for Integrative Neuroscience, Purdue University, West-Lafayette, IN, USA.\\ 
$^{4}$Department of Basic Medical Sciences, Purdue University, West-Lafayette, IN, USA.\\
$^{5}$Weldon School of Biomedical Engineering, Purdue University, West-Lafayette, IN, USA.\\
\bigskip
* jgonicor@purdue.edu

\end{flushleft}


\begin{abstract}
Multi-site studies are becoming important to increase statistical power, enhance generalizability, and to improve the likelihood of pooling relevant subgroups together activities. Even with harmonized imaging sequences, site-dependent variability can mask the advantages of these multi-site studies. The aim of this study was to assess multi-site reproducibility in resting-state functional connectivity fingerprints, and to improve identifiability of functional connectomes. The individual fingerprinting of functional connectivity profiles is promising due to its potential as a robust neuroimaging biomarker. We evaluated, on two independent multi-site datasets, individual fingerprints in test-retest visit pairs within and across two sites and present a generalized framework based on principal component analysis to improve identifiability. Those components that maximized differential identifiability of a training dataset were used as an orthogonal connectivity basis to reconstruct the functional connectomes of training and validation sets. The optimally reconstructed functional connectomes showed a substantial improvement in individual fingerprinting within and across the two sites relative to the original data. A notable increase in ICC values for functional edges and resting-state networks was also observed. Improvements in identifiability were not found to be affected by global signal regression. Post-hoc analyses assessed the effect of the number of fMRI volumes on identifiability and showed that multi-site differential identifiability was for all cases maximized after optimal reconstruction. The generalizability of the optimal set of orthogonal basis of each dataset was evaluated through a leave-one-out procedure. Overall, results demonstrate that the framework presented in this study systematically improves identifiability in resting-state functional connectomes in multi-site studies.
\end{abstract}


\section{Introduction}
\label{Intro}

Multi-site functional magnetic resonance imaging (fMRI) studies are increasingly important for understanding the structure and function of a healthy brain and also subsequent to neuropathology. Recent examples of multi-site imaging initiatives include The Human Connectome Project \cite{VanEssen2012ThePerspective,VanEssen2013TheOverview}, the 1,000 Functional Connectomes Project (\url{http://fcon_1000.projects.nitrc.org}), and disease-oriented initiatives such as the Functional Biomedical Informatics Research Network for schizophrenia \cite{Keator2016TheRepository} and the Alzheimer's Disease Neuroimaging Initiative \cite{Jack2008TheMethods}, among others \cite{VanHorn2009MultisiteTrials.}. Multi-site studies achieve larger sample sizes by including cohorts recruited at the different sites. On one hand this allows for higher statistical power and better generalization of the results than may be achieved with potentially limited availability of patients or funds at a single site. On the other hand proper assessment of these data requires principled methodologies, including multivariate analyses coupled with cross-validation designs \cite{Jack2008TheMethods,VanHorn2009MultisiteTrials.,Friedman2006ReducingDifferences,Mulkern2008EstablishmentConsortium,Brown2011MultisiteData}. Known challenges in multi-site acquisitions and their subsequent analyses include the scanner-dependent variability that can mask true underlying changes in brain structure and function. Even when using identical (let alone ``comparable'') imaging sequences and parameters, potential site-dependent differences might arise due to a range of physical variables, including  field inhomogeneities, transmit and receive coil configurations, system stability, system maintenance, scanner drift over time and many others \cite{VanHorn2009MultisiteTrials.,Friedman2006ReducingDifferences,Voyvodic2006ActivationStrengths}. Determining and minimizing these unwanted site-dependent variations have become critical elements in the design of multi-site fMRI studies.    

Many studies have investigated the variation and stability of simple behavioral, motor or memory tasks in multiple sites using fMRI. Such  studies have typically used ANOVA models or variance component analysis to examine the variability and extent of overlap of activation maps in task-based fMRI scans acquired across multiple sites \cite{Brown2011MultisiteData,Voyvodic2006ActivationStrengths,Friedman2008Test-retestStudy,Casey1998ReproductibilityTask,Gountouna2010FunctionalTask,Suckling2008ComponentsPower,Yendiki2010Multi-siteIndices,Zou2005ReproducibilityNetwork}. In contrast, only a few studies have assessed the variations in resting-state fMRI across sites \cite{Noble2017MultisiteConnectivity}. These studies used variance component analysis, intra class correlation (ICC) coefficient and/or coefficient of variance to evaluate inter-site and inter-subject variability in connectivity scores, cluster size and temporal signal-to-noise ratio in regions of interest for default mode networks derived from seed-based or independent component analysis \cite{Deprez2018Multi-centerMeasures,Noble2017MultisiteConnectivity,Noble2017InfluencesUtility,Jovicich2016LongitudinalStudy,Marchitelli2016Test-retestTechniques,Huang2012ReproducibilityScanners,Turner2013ASchizophrenia,Feis2015ICA-basedFMRI,Jann2015FunctionalNetworks}.

Resting-state fMRI (rs-fMRI) measures the spontaneous neural activity in the brain and determines the default functional connectivity between brain regions. rs-fMRI has gained wide-spread attention and is used to investigate brain functional connectivity in the normal healthy brain \cite{Biswal1995FunctionalMri,Fox2007SpontaneousImaging,Greicius2003FunctionalHypothesis.,Beckmann2005InvestigationsAnalysis.,Shehzad2009TheReliable} as well as in many clinical populations \cite{Mayer2011FunctionalInjury,Broyd2009Default-modeReview,Contreras2015TheAdults}. In recent years, the research areas of network neuroscience and brain connectomics have become central to the understanding of the human brain as a network. In consequence, graph theory and network science methods have been widely used to investigate functional connectivity \cite{Bullmore2009ComplexSystems,Rubinov2010ComplexInterpretations,Sporns2014ContributionsNeuroscience,He2009NeuronalDisease,Fornito2016FundamentalsAnalysis,Braun2012TestretestMeasures,Wang2011GraphData}. A functional connectome (FC) is a symmetric square matrix that estimates the level of functional coupling between pairs of brain regions. Each entry is the correlation between the blood oxygenation level dependent (BOLD) signals observed in two different brain regions. Various graph theoretical measures may be used to investigate FC networks \cite{Fornito2016FundamentalsAnalysis}. 

One important avenue of investigation is to explore differences in FC profiles at an individual, rather than group, level \cite{Schultz2016HigherReconfiguration}. Group averages represent robust connectivity patterns, but inherently mask subject-specific features. Differences in FC profiles in individuals, relative to the group level, have been demonstrated \cite{Mira-Dominguez2014Connectotyping:Connectome,Finn2015FunctionalConnectivity,Finn2017CanConnectivity,Gordon2017Individual-specificCorrelations,Waller2017EvaluatingFingerprints,Gordon2017PrecisionBrains,Zimmermann2018Subject-SpecificityConnectivity,Satterthwaite2018PersonalizedNetworks,Greene2018Task-inducedTraits,Gonzalez-Castillo2017Task-basedQuestions,Amico2018TheConnectomes,Pallares2018ExtractingConnectivity,Svaldi2018TowardsConnectomes} and may help in developing robust neuroimaging-based biomarkers, or even for making subject-level inferences. Robust individual differences in functional connectivity have been termed ``fingerprints'', and may be demonstrated by the self-identification of subjects by correlating test and retest visits over a body of subjects \cite{Finn2015FunctionalConnectivity,Amico2018TheConnectomes,Vanderwal2017IndividualConditions,Shen2017UsingConnectivity,Yoo2018Connectome-basedDatasets,Shah2016ReliabilityState}. Fingerprinting relies on the fact that subjects are expected to exhibit an inter-session variability that is less than the inter-subject variability (i.e., they resemble themselves across visits more strongly than they resemble other subjects). The ability to pair the FCs coming from the same subject reflects the inherent level of identifiability of the connectivity dataset. 

This study explores the question of variability in the identifiability of subjects in a multi-site scenario, providing a framework to minimize the unwanted site-dependent variations and enhance identifiability on functional connectomes. To do so, we evaluated two independent multi-site resting-state fMRI datasets. To date, identifiability has been studied where test and retest rs-fMRI scans have been conducted in the strictly controlled scenario involving  the same scanner, the same imaging sequences, same-day image acquisitions, and constant processing over all data \cite{Finn2015FunctionalConnectivity,Amico2018TheConnectomes,Yoo2018Connectome-basedDatasets}. For example, Amico et al., \cite{Amico2018TheConnectomes} determined the identifiability of subjects based on test-retest visits on one site. Herein we extend the investigation of identifiability by relaxing a number of these conditions. In particular, one dataset (Purdue) used two different scanners and varied in the number of days between visits, whereas the second dataset (Yale) involved two identically configured imaging-sites. For all cases, the impact of global signal regression as part of the data processing pipeline was also assessed.
We optimally reconstructed the FCs using those principal components that maximized multi-site identifiability across all visits, thereby and serving as an orthogonal basis for the functional connectivity. This was performed both with and without global signal regression. For each of these cases, we then compared the multi-site identifiability obtained from the original and optimal reconstructed 
FCs. In all cases, the reconstruction process produced significantly enhanced identifiability across imaging systems, providing strong motivation for application of this approach to increase the statistical power and generalizability of results for multi-site fMRI studies.

\section{Methods}
\label{methods}
\subsection{Participants}
\subsubsection{Purdue dataset}
A cohort consisting of 23 undergraduate and graduate students (12 male and 11 female; ages 18-28 years) participated in a total of four imaging sessions (0-21 days apart) at two sites. None of the participants reported any history of neurological disorders. At \textit{site1} two imaging sessions were conducted using a 3T General Electric Signa HDx and a 16-channel brain array (Nova Medical). At \textit{site2} two imaging sessions were conducted using a 3T GE Discovery MR750 and a 32-channel brain array (Nova Medical). The two imaging sessions at a given site were conducted on the same day (i.e., 0 days apart).

\subsubsection{Yale dataset}
An open source dataset available at \url{ http://fcon_1000.projects.nitrc.org/indi/retro/yale_trt.html} consisting of 12 (six male and six female; ages 27-56 years) participants was used in the study. The subjects participated in a total of four imaging sessions at two sites approximately one week apart. Data were acquired on two identically configured Siemens 3T Tim Trio scanners at Yale University using a 32-channel head coil \cite{Noble2017InfluencesUtility}.

\subsection{MRI Data Acquisition} 
\subsubsection{Purdue dataset}
Each imaging session (independent of site) consisted of a structural T$_{1}$ weighted scan and two rs-fMRI scans (\textit{test} and \textit{retest}, eyes open and 9 min and 48 sec). The high-resolution T$_{1}$ scan used for registration and segmentation purposes consisted of 3D fast spoiled gradient recalled echo sequence: TR/TE = 5.7/1.976 msec; flip angle = $73^{\circ}$; 1 mm isotropic resolution and the rs-fMRI scans with common imaging parameters across sites consisted of blipped echo-planar imaging: TR/TE = 2,000/26 msec; flip angle = $35^{\circ}$; 34 slices; acceleration factor = 2; Field of View = 20 cm; voxel size = 3.125 x 3.125 x 3.80 mm and 294 volumes.

Note that eight rs-fMRI scans were conducted in total on each subject (184 total scans; see Figure \ref{scans}) and divided into a \textit{training} and a \textit{validation} sets. The two runs acquired in the first session at each of the two sites (four total) were incorporated into the \textit{training} set. Similarly the remaining four rs-fMRI scans, those corresponding to the two runs acquired in the second imaging session at that each site, were incorporated into the \textit{validation} set. 

\subsubsection{Yale dataset}
Each imaging session (independent of site) consisted of a structural T$_{1}$ weighted scan and six rs-fMRI scans (eyes open and 6 min). 
T$_{1}$-weighted 3D anatomical scans were acquired using a magnetization prepared rapid gradient echo (MPRAGE) sequence: TR/TE = 2400/1.18 msec; flip angle = $8^{\circ}$; 1 mm isotropic resolution
and the rs-fMRI scans with multiband echo-planar imaging: TR/TE = 1000/30 msec; flip angle=$55^{\circ}$; 75 slices; acceleration factor = 5; Field of View = 22 cm; voxel size = 2 x 2 x 2 mm and 360 volumes \cite{Noble2017InfluencesUtility}.

Two of the six rs-fMRI scans from each imaging session were used as the \textit{test} and \textit{retest}. The imaging sessions were divided in to \textit{training} and \textit{validation} sets in the same way as Figure  \ref{scans}.

\subsection{Data Processing}
Both Purdue and Yale datasets were processed with the same processing pipeline, as described below. 

rs-fMRI data were processed using functions from AFNI \cite{Cox1996AFNI:Neuroimages} and FSL \cite{Jenkinson2012FSL,Smith2004AdvancesFSL} using in-house MATLAB code following steps from \cite{Amico2017MappingConsciousness}. Structural T$_{1}$ images were first denoised using the filters described in \cite{Coupe2010RobustImages.,Coupe2008AnImages,Wiest-Daessle2008RicianDT-MRI} (using FSL \textit{fsl\_anat}) to improve signal-to-noise ratio and effect bias-correction. Images also underwent intensity normalization (AFNI \textit{3dUnifize}). Structural images were then segmented (FSL FAST) into gray matter (GM), white matter (WM) and cerebrospinal fluid (CSF) tissue masks. 

rs-fMRI BOLD timeseries were processed in the subject's native space. The first four volumes were discarded to remove spin history effects, leaving 290 volumes for processing. The 4D BOLD timeseries was then passed through outlier detection (AFNI \textit{3dToutcount}), despiking (AFNI \textit{3dDespike}), slice timing correction (AFNI \textit{3dTshift}), and subsequently underwent volume registration (AFNI \textit{3dvolreg}) to the minimized bounding volume. The rs-fMRI BOLD timeseries were then aligned to the T$_{1}$ structural scan (AFNI \textit{align\_epi\_anat.py}). Voxel-wise spatial smoothing was applied independently within each of the GM, WM and CSF masks, using a 4mm full-width-at-half-maximum isotropic Gaussian Kernel (AFNI \textit{3dBlurinMask}). The resulting BOLD timeseries were then scaled to a maximum (absolute value) of 200, and data were censored to remove outlier timepoints. Censoring of individual rs-fMRI volumes occurred if the motion derivatives had a Euclidean norm \cite{Jones2010SourcesDisorder} above 0.4. Censoring involved removal not only of the volume at which this high norm was observed, but also the immediately preceding and following volumes, given that effects of motion may be carried across timepoints. Entire rs-fMRI timeseries were discarded if more than 100 volumes (34\% of the volumes) were censored. Only the subjects for which all eight rs-fMRI scans survived motion censoring were included in the analysis. 

Purdue dataset: Out of $23$ subjects, a final pool of $18$ subjects (144 rs-fMRI scans) was retained for analysis. Three of the original 184 rs-fMRI scans---and their associated three subjects---were rejected due to excessive motion. An additional two subjects were rejected due to poor registration to the template of at least one of the sessions. Yale dataset: $11$ out of $12$ subjects were included. One subject was dropped after failure in the NIFTI reconstruction of the raw DICOM images.

To assess the impact of global signal regression on the reconstruction procedure and subsequently identifiability, all included fMRI runs were evaluated both after being detrended with (\textit{GSR}) and without (\textit{NoGSR}) global signal regression. 
Each detrending (AFNI \textit{3dDeconvolve}) approach incorporated the following common regressors: 
(1) very low frequency fluctuations as derived from a bandpass [0.002-0.01Hz] filter (AFNI \textit{1dBport}); (2) the 12 motion parameters, consisting of three linear translations [x,y,z], three rotations [pitch, yaw, roll] and the corresponding set of  first derivatives \cite{Power2012SpuriousMotion,Power2014MethodsFMRI}; and (3) the voxel-wise local neighborhood (40mm) mean WM timeseries (AFNI \textit{3dTproject}) \cite{Jo2010MappingRemoval.}.  The data at this point represent the \textit{NoGSR} dataset.  Incorporation of a fourth regressor source---the whole-brain mean GM timeseries---in the detrending stage results in the \textit{GSR} dataset.

For connectivity analysis on a regional basis, the grey matter brain atlas from \cite{Shen2013GroupwiseIdentification} was warped to each subject's native space by linear and non-linear registration (AFNI \textit{auto\_warp.py} and \textit{3dAllineate}). This brain parcellation consists of $278$ regions of interest (ROIs). Note that data from the cerebellum (including a total of 30 ROIs) were discarded, because the acquired data did not completely cover this structure for all subjects. This resulted in a final GM partition of $248$ ROIs.

A functional connectivity matrix (namely the functional connectome; FC) was computed for each rs-fMRI scan through correlation of the mean time series from each of the 248 ROIs (\textit{MATLAB} command \textit{corr}). The resulting  square, symmetric FC matrices were not thresholded or binarized. Each FC matrix was ordered into seven cortical sub-networks, as proposed by Yeo et al. \cite{ThomasYeo2011TheConnectivity}, and an additional eighth sub-network comprising sub-cortical regions was added \cite{Amico2017MappingConsciousness}.For each dataset (Purdue and Yale), this resulted in eight functional connectomes per subject (four from each site; two \textit{training} and two \textit{validation}).

\subsection{Differential Identifiability extended for Multi-Site studies}

The upper triangular of each FC (test and retest) for the \textit{training} data was vectorized and added to a matrix where the columns were runs and the rows represent the functional connectivity patterns. Hence, this matrix had $\binom{248}{2}$ rows and N*4 columns ($4$ runs per subject; N subjects). Principal component analysis (PCA) was used to extract M = N*4 principal components (i.e., functional connectivity eigenmodes) from the vectorized \textit{training} dataset (\textit{MATLAB} command \textit{pca}). The principal components (PCs) were arranged in descending order of their explained variance. These PCs were then projected back into each subject's FC space to obtain individual reconstructed functional connectomes as analogously done by Amico et al. \cite{Amico2018TheConnectomes}. Below we extend this approach for multi-site acquisitions.

For individual fingerprints of subjects within and across sites, the identifiability matrix (\textbf{I}) was created by correlating the subjects’ test and retest FCs within and across the two sites. This gave rise to a multi-site identifiability matrix, \textbf{I} which consisted of Pearson's correlation coefficients. For the particular case of two imaging sites, the test-retest combinations created four blocks ($\textbf{I}^{ij}$) in the identifiability matrix \textbf{I}, 
\begin{align*}
\textbf{I} &= \begin{bmatrix}
\textbf{I}^{11} & \textbf{I}^{12} \\
\textbf{I}^{21} & \textbf{I}^{22}
\end{bmatrix}
\end{align*}
where $\textbf{I}^{ij}$ contained Pearson's correlation coefficient obtained by correlating FCs from the site\textit{i} test session with the FCs from the site\textit{j} retest session. $\textbf{I}^{11}$ and $\textbf{I}^{22}$ represent the fingerprinting of the subjects within the two sites and $\textbf{I}^{12}$ and $\textbf{I}^{21}$ represent the fingerprinting of the subjects across the two sites. 

For each test-retest [site\textit{i}, site\textit{j}] pair, differential identifiability ($<I_{diff}^{ij}>$) was calculated from the block $\textbf{I}^{ij}$ following the procedure from \cite{Amico2018TheConnectomes} 
\begin{equation*}
<I^{ij}_{diff}> \ =\  <I^{ij}_{self}> - <I^{ij}_{others}> 
\end{equation*}
where 
\begin{align*}
<I^{ij}_{self}> & \  =\ \frac{1}{N} \sum_{k=1}^{N} I^{ij}_{self} (k) \\\\
I^{ij}_{self} (k) &\  =\ I^{ij}_{kk} \ , \ \ \ \  \forall \ k=1,2,\ldots,N
\end{align*}
where N is the number of subjects (N=18 for Purdue dataset; N=11 for Yale dataset).

$\textbf{I}^{ij}_{self}$, defined as \textit{self identifiability}, is a vector of length N and contains the main diagonal elements $I_{self}^{ij}(k)$ of the block $\textbf{I}^{ij}$, and denotes the correlation between the repeat visits of the same subject. The average of the main diagonal elements for the block $\textbf{I}^{ij}$, $<I^{ij}_{self}>$, represents the overall self correlation for the [site\textit{i}, site\textit{j}] pair. 

\begin{align*}
<I^{ij}_{others}> & \   =\  \frac{1}{N} \sum_{k=1}^{N} I^{ij}_{others} (k) \\\\
I^{ij}_{others} (k) &\  =\  \frac{1}{2} \bigg( \frac{1}{N-1} \sum_{l=1}^{N} I^{ij}_{kl} + \frac{1}{N-1} \sum_{l=1}^{N} I^{ij}_{lk}  \bigg)  \ ,  \ \forall \ l\neq k
\end{align*}

For the \textit{k-th} subject $I^{ij}_{others} (k)$ is an element of the vector $\textbf{I}^{ij}_{others}$ and is obtained by the average of the \textit{k-th} row and \textit{k-th} column, excluding the main diagonal entry of the block $\textbf{I}^{ij}$, and defines the average correlation of the \textit{k-th} subject's FCs (test and retest) with all other subjects. 
$<I^{ij}_{others}>$ is the average of all $I^{ij}_{others} (k)$ of the block $\textbf{I}^{ij}$, and defines an overall mean correlation between visits of different subjects for the [site\textit{i}, site\textit{j}] pair. 

For visits associated with the [site\textit{i}, site\textit{j}] pair, $<I_{diff}^{ij}>$ characterizes the difference between the average within-subject FC similarity and the average between-subject FC similarity. The higher the value of $<I_{diff}^{ij}>$, the stronger is the overall fingerprinting of the population for the [site\textit{i}, site\textit{j}] pair. 

To maximize the fingerprinting of the population across all the [site\textit{i}, site\textit{j}] visit pairs, the average of the four $<I_{diff}^{ij}>$ values was used, where
\begin{equation*}
<<I_{diff}>>\  =\  \frac{1}{n} \sum_{i=1}^{n}\sum_{j=1}^{n} <I_{diff}^{ij}>
\end{equation*}
Here, n=2 is the number of sites for both Purdue and Yale datasets.

Multi-site differential identifiability $<<I_{diff}>>$ is then maximized by the selection of subsets of \textbf{m} PCs from the total number (M = N*4) of PCs obtained from the training set.  For each subset of the first \textbf{m} PCs, the subjects’ test-retest FCs were reconstructed, and $<<I_{diff}>>$ was calculated from these data. The optimal number of PCs, \textbf{m*}, maximizes the value of $<<I_{diff}>>$, namely $<<I_{diff}^{*}>>$, as given by \cite{Amico2018TheConnectomes}: 
\begin{equation*}
<<I_{diff}^{*}>>\ =\ argmax_{m \in M} \ <<I_{diff}>> (\textit{m})
\end{equation*}

The \textbf{m*} PCs were used to reconstruct the individual FCs (for both visits---i.e., test and retest) for the \textit{training} and \textit{validation} sets. The identifiability matrices computed from the original and reconstructed data for each of the \textit{training} and \textit{validation} sets were then compared.

Analogously, when focused on a particular [site\textit{i}, site\textit{j}] visit pair, we may obtain $m^{ij*}$ as 
\begin{equation*}
<I_{diff}^{ij\ *}>\ =\ argmax_{m^{ij} \in M} \ <I_{diff}^{ij}> (\textit{m})
\end{equation*}

\subsection{Statistical Analysis}

Differential Identifiability ($\textbf{I}_{diff}^{ij} $) was computed for each [site\textit{i}, site\textit{j}] pair from $\textbf{I}^{ij}$ as follows
\begin{equation*}
\textbf{I}_{diff}^{ij}= \textbf{I}_{self}^{ij} - \textbf{I}_{others}^{ij}
\end{equation*}

For the \textit{k-th} subject the value of $I_{diff}^{ij} (k) $ was calculated as 
\begin{equation*}
I_{diff}^{ij} (k)= I_{self}^{ij}(k) - I_{others}^{ij}(k)
\end{equation*}

$I_{diff}^{ij}(k)$ characterizes the differential identifiability on a subject level and quantifies the difference between the \textit{k-th} subject's FC \textit{self identifiability} and its similarity with other subjects' functional connectomes. The higher the value of $I_{diff}^{ij}(k)$, the higher is the identifiability of the \textit{k-th} subject among the cohort. 

Pairwise comparisons were done on the distributions of $\textbf{I}_{diff}^{ij}$ obtained from the original and reconstructed data, for both the \textit{training} and \textit{validation} sets, using the Wilcoxon signed rank test followed by a Bonferroni correction on each subset of tests (e.g., four tests were conducted on the each of the \textit{NoGSR} and \textit{GSR} \textit{training} and \textit{validation} sets, so a correction for four tests was performed). All such analyses were conducted in R \cite{Team2013R:Computing}. Any pairwise comparison was considered significant if $p_{Bonferroni} <$ 0.05. Similar comparisons were also made between the distributions of $\textbf{I}_{diff}^{ij}$ as obtained from reconstructions for original data with (\textit{GSR}) and without (\textit{NoGSR}) global signal regression.

The intraclass correlation coefficient (ICC) was used to assess the agreement of an edge (functional connectivity value between two brain regions) between visits of subjects on each [site\textit{i}, site\textit{j}] pair. ICC \cite{Shrout1979IntraclassReliability.,McGraw1996FormingCoefficients.} is generally used to assess the agreement between measurements for different groups. The stronger the resemblance between the measurements, the higher is the ICC value. Furthermore, a bootstrap procedure was applied when computing ICC to avoid biases induced by a small subset of the population. In each of $100$ iterations 75\% of the population was selected at random, and the ICC was calculated for each edge. The averages over all iterations were used to compare the edgewise ICC values of the original and the reconstructed data. ICC values for the resting-state functional networks of \cite{Yeo2011TheConnectivity}, for both the original and reconstructed data, were computed by averaging over the ICC values for the edges that belonged to each functional network. Using the aforementioned bootstrap procedure, edgewise ICC was also computed from all $4$ visits across the two sites and these edgewise ICC were averaged over each brain region from \cite{Shen2013GroupwiseIdentification} to compare the reproducibility, between training and validation sets, of connectivity in each brain region across the original and reconstructed data. This entire edgewise ICC procedure was repeated for each of the \textit{GSR} and \textit{NoGSR} modalities. 
\begin{figure}
\centering
\includegraphics[scale =0.5,trim= {3cm 5.5cm 2.5cm 6.5cm}, clip=true]{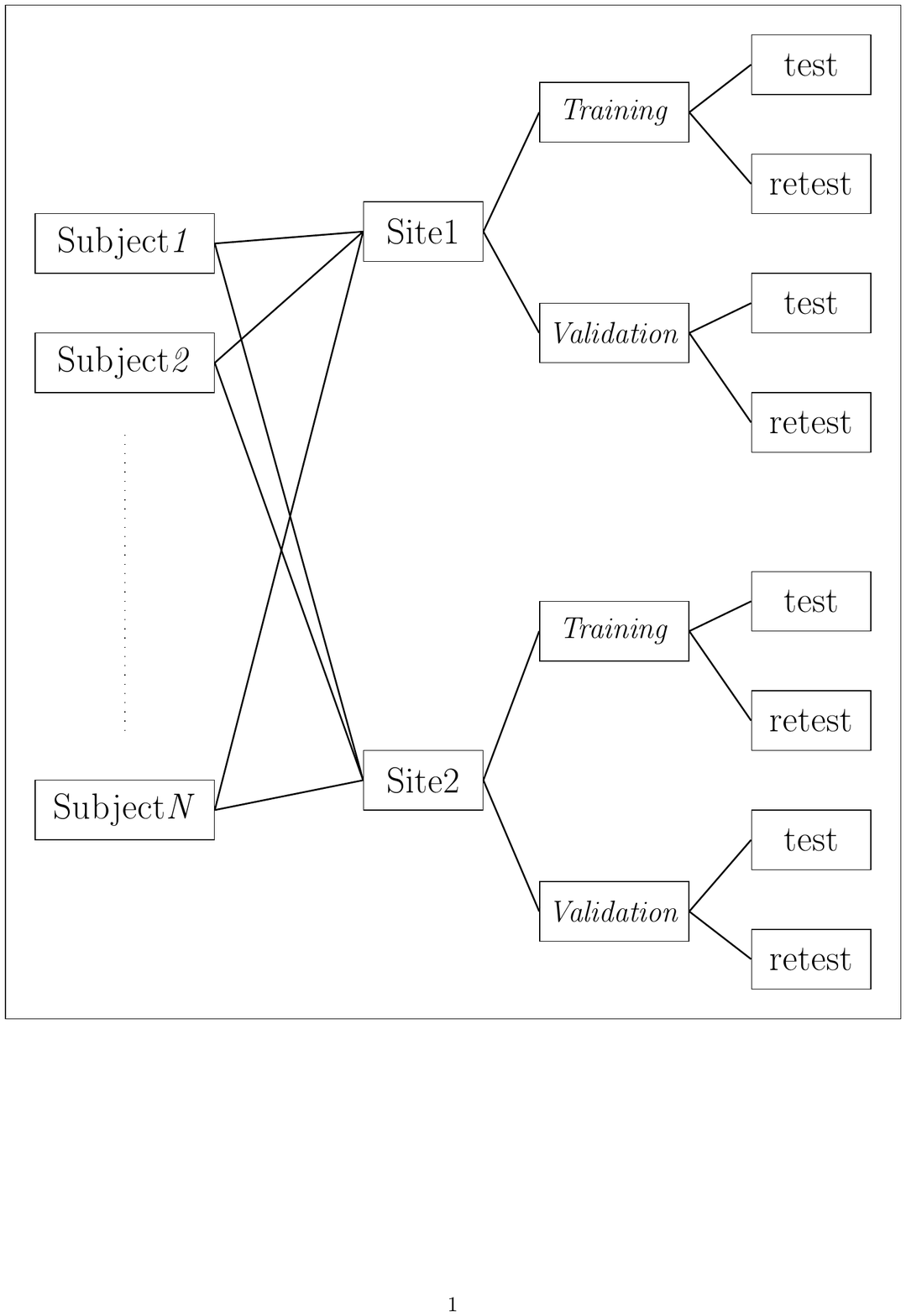}
\caption{Diagram of the resting-state fMRI acquisitions for both datasets. Subjects  underwent two imaging sessions (\textit{Training} and \textit{Validation}) at each of two MRI sites (Site1 and Site2), wherein each session comprised two resting-state runs (test and retest). After quality checks, the Purdue dataset included 18 subjects and the Yale dataset included 11 subjects. This setup produced a total of eight runs and associated functional connectomes (FC) per subject.}
\label{scans}
\end{figure}

\section{Results}
\label{results}

The Purdue dataset used for this study consisted of two fMRI sessions (each session consisted of test and retest pair of rs-fMRI scans) per subject on two different sites. After quality checks, 18 subjects with eight FCs per subject were used for Purdue dataset (see Methods). Building upon \cite{Amico2018TheConnectomes}, we here expanded the concept of identifiability for multiple acquisitions on multiple sites. We evaluated this method by splitting our dataset (see Figure \ref{scans}) into \textit{training} and \textit{validation} sets. The \textit{training} dataset consisted of four FCs per subject (test-retest at site1, test-retest at site2). Analogously, the \textit{validation} dataset consisted of another four FCs per subject (for the same subjects as the \textit{training} dataset; test-retest at site1, test-retest at site2). 

\begin{figure}
\centering
\includegraphics[scale =0.9,trim= {3cm 8cm 2cm 9cm}, clip=true]{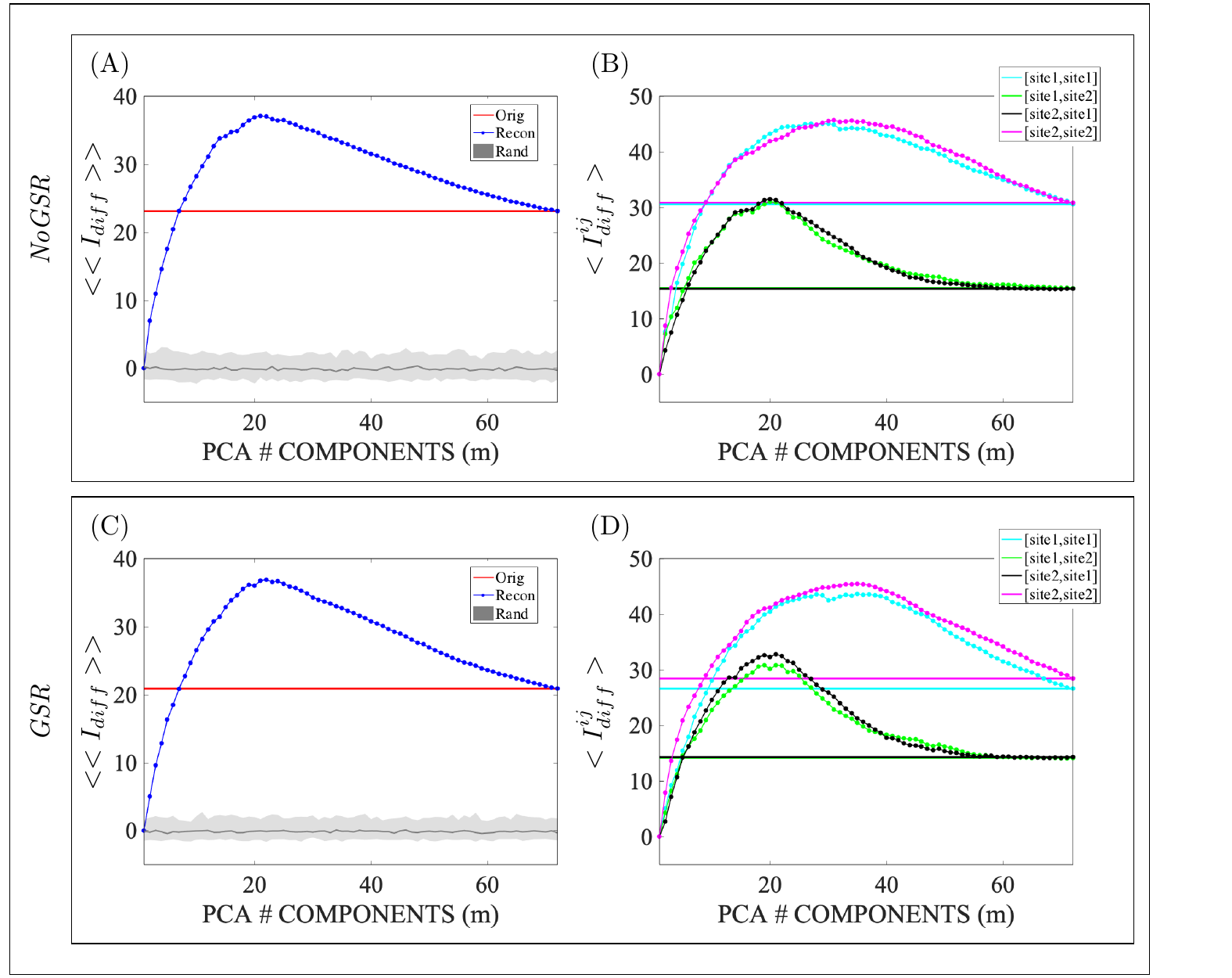}
\caption{Purdue dataset. Multi-site differential identifiability $(<<I_{diff}>>*100)$ and differential identifiability of each [site\textit{i}, site\textit{j}] pair, $(<I_{diff}^{ij}>*100)$ for training data as a function of the number of principal components (PCs) used for reconstruction for resting-state data without global signal regression (\textit{NoGSR}; (A) and (B)); and with global signal regression (\textit{GSR}; (C) and (D)). In all figures solid lines denote $<<I_{diff}>>$ and $<I_{diff}^{ij}>$ as computed from the original FCs, whereas lines with circles denote the differential identifiability for reconstructed FCs as a function of \textbf{m}, the included number of components. In (A) and (C), the gray (shaded) area denotes the 95\% confidence interval for $<<I_{diff}>>$ over 100 random permutations of the test-retest FC pairs at each value of \textbf{m}. 
It may be observed that the benefit of reconstruction on differential identifiability was not dependent on the exclusion/inclusion of global signal regression.}
\label{Idiff}
\end{figure}

When assessing the Purdue \textit{training} dataset, FCs were decomposed and subsequently reconstructed based on PCA by using each subset of first \textbf{m} number of components out of the total (M = 72). For each number of PCA components \textbf{m}, $<<I_{diff}>>$ was computed from the reconstructed data (see Methods) and compared to $<<I_{diff}>>$ obtained from original data.
Figure \ref{Idiff} shows $<<I_{diff}>>$ computed from the original and, iteratively, from the reconstructed data as a function of (\textbf{m}), the number of PCs preserved. $<<I_{diff}>>$ peaked at \textbf{m*} = 21 for \textit{NoGSR} and \textbf{m*} = 22 for \textit{GSR} datasets. These \textbf{m*} PCs extracted from the \textit{training} set were used as a fixed orthogonal connectivity basis (i.e. PCA loadings) to reconstruct the functional connectomes (denoted by Recon) of the \textit{training} and \textit{validation} sets for comparing identifiability obtained from the original FCs (Orig). 

When looking at $<I_{diff}^{ij}>$ for different [site\textit{i}, site\textit{j}] visit pairs for \textit{NoGSR} and \textit{GSR}, we found different optimal numbers of components ($m^{ij}$). Within-site configurations peaked at $29$ and $31$ components respectively (\textit{NoGSR}) and at $35$ components (\textit{GSR}). Between-sites configurations both peaked at $20$ components (\textit{NoGSR}) and at $21$ components ((\textit{GSR})). Briefly, more components were included in the optimal reconstruction (and hence more variance was preserved) for within-site configurations whereas less components were included for between-site configurations (and hence less variance was preserved) for optimal identifiability. A summary of $<I_{diff}^{ij}>$ and the corresponding $m^{ij*}$ for all configurations is shown in Table \ref{I_diff_ij}. 

\begin{figure}
\centering
\includegraphics[scale =0.9,trim= {2.8cm 9cm 2cm 10cm}, clip=true]{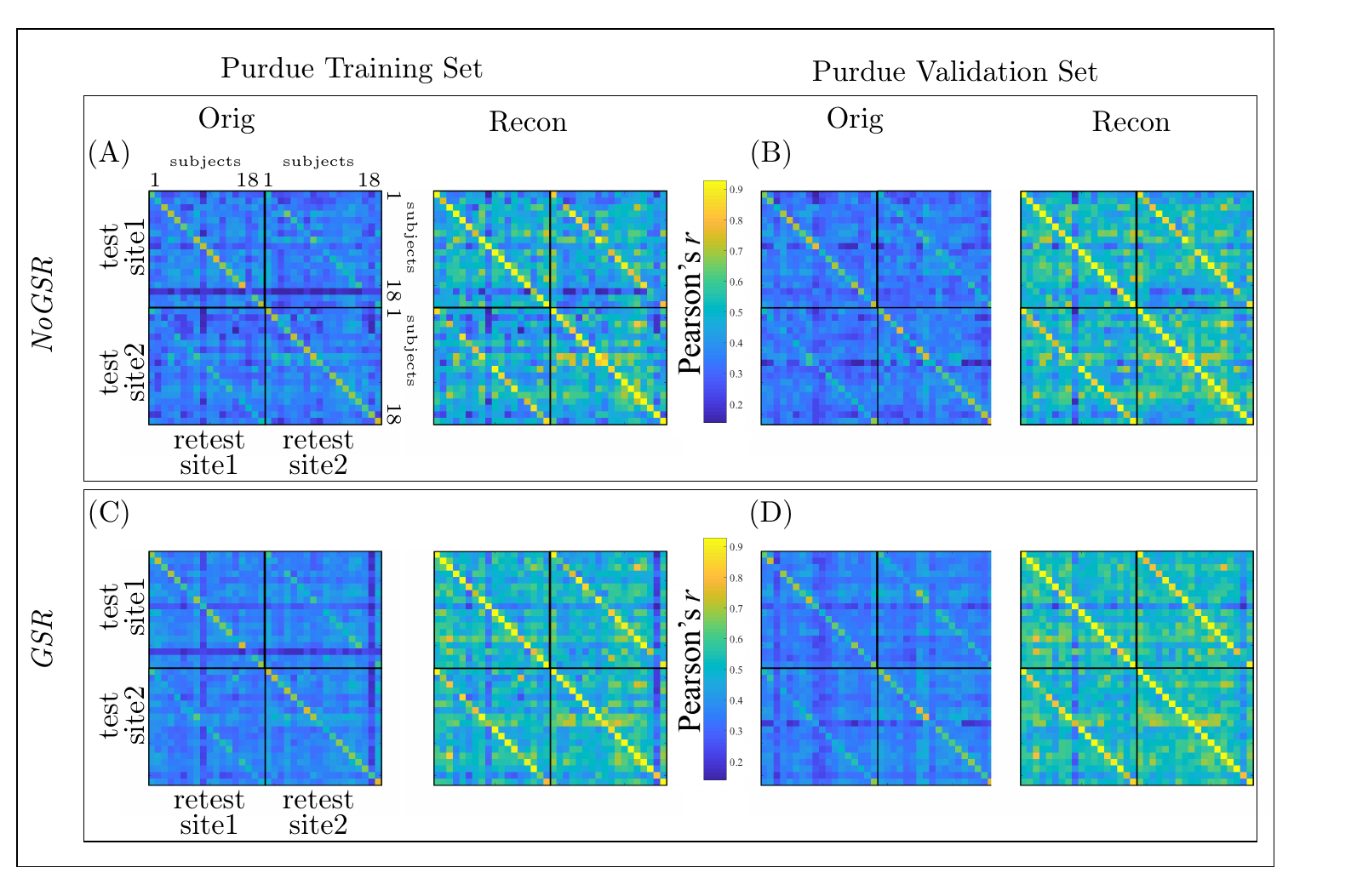}
\caption{Purdue dataset. Identifiability matrices (\textbf{I}) of the original (Orig) and reconstructed (Recon) data for the \textit{Training}, (A) and (C), and \textit{Validation}, (B) and (D) sets of resting-state functional connectomes without global signal regression (\textit{NoGSR}; (A) and (B)) and with global signal regression (\textit{GSR}; (C) and (D)). The Identifiability matrix (\textbf{I}) has a blockwise structure where each block is $I^{ij}$, representing the identifiability for the [site\textit{i}, site\textit{j}] pair.
Note that identifiability was meaningfully improved across sites regardless of the exclusion/inclusion of global signal regression.}
\label{Ident_matrices}
\end{figure}

Identifiability matrices (\textbf{I}) consisting of Pearson’s correlation coefficient between FCs of subjects' test and retest visits across and within the two sites were computed, expanding on \cite{Amico2018TheConnectomes}. The identifiability matrices obtained from reconstructed FCs using \textbf{m*} PCs were compared to the ones obtained from original data. 
Figure \ref{Ident_matrices} illustrates that the identifiability matrices obtained from optimally reconstructed functional connectomes outperformed the original FCs. The individual fingerprint of the subjects (main diagonal of each block $\textbf{I}^{ij}$) within and across the sites were always higher at the optimal reconstruction for both \textit{NoGSR} and \textit{GSR} datasets. 

\begin{figure}
\centering
\includegraphics[scale =0.9,trim= {3cm 8cm 2cm 9cm}, clip=true]{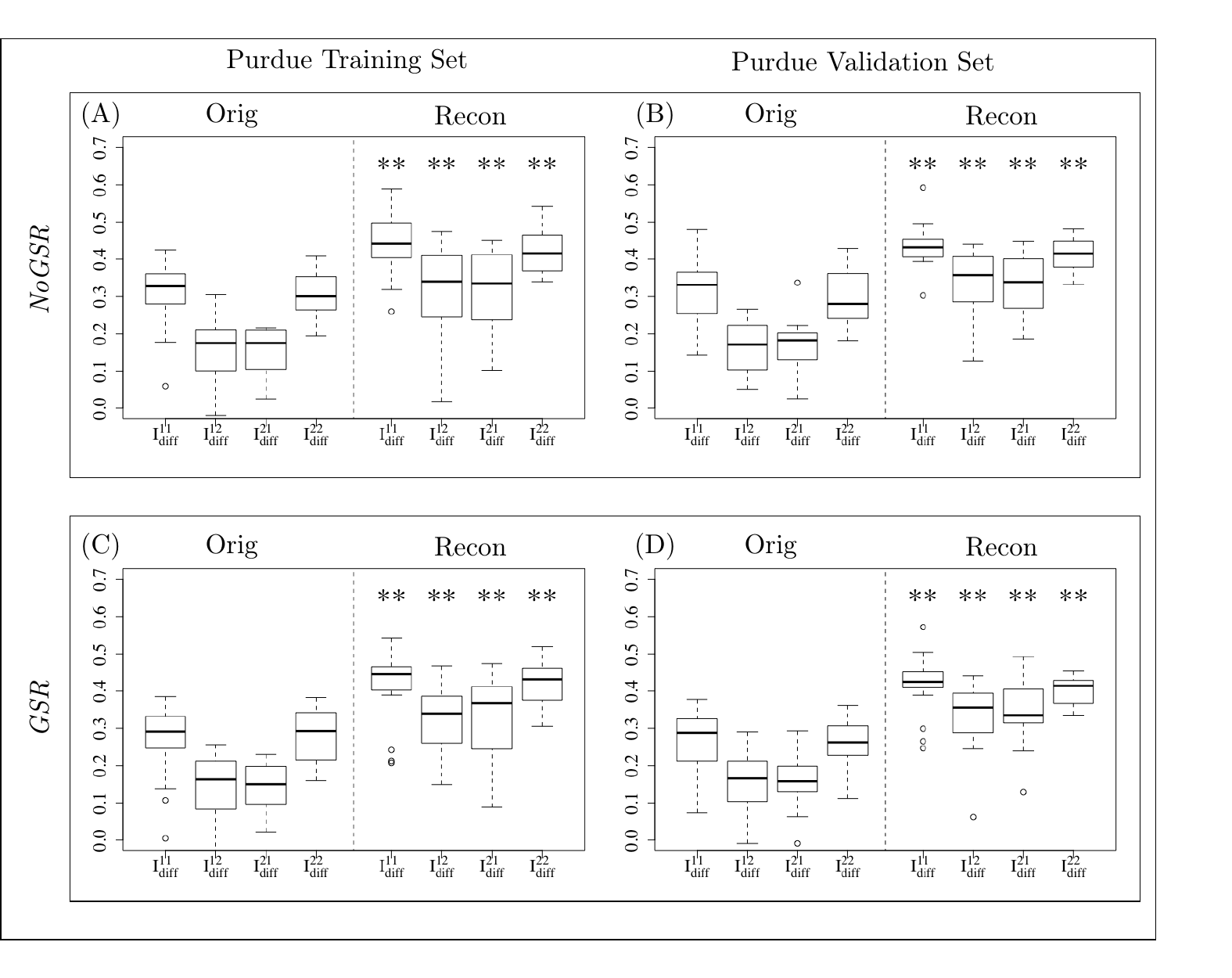}
\caption{Purdue dataset. Box plots of Differential Identifiability ($\textbf{I}_{diff}^{ij}$) computed from each block of the Identifiability matrix (i.e., $\textbf{I}^{ij}$) for the original (Orig) and optimally reconstructed (Recon) data without global signal regression (\textit{NoGSR}; (A) and (B)) and with global signal regression (\textit{GSR}; (C) and (D)). Values of Pearson's \textit{r} that are significantly higher ($p_{Bonferroni}$ < 0.05, Wilcoxon signed rank) for Recon relative to Orig are marked by double asterisks.
Note that distributions of $\textbf{I}_{diff}^{ij}$ were found to be unaffected by exclusion/inclusion of global signal regression.}
\label{Icont}
\end{figure}

Differential Identifiability ($\textbf{I}_{diff}^{ij})$) for each [site\textit{i}, site\textit{j}] pair was computed from $\textbf{I}^{ij}$ blocks (see Methods). The distributions of $\textbf{I}_{diff}^{ij}$ obtained from original and optimally reconstructed data were compared. 
Figure \ref{Icont} shows that the distributions of $\textbf{I}_{diff}^{ij}$ for each [site\textit{i}, site\textit{j}] pair was significantly higher ($p_{Bonferroni} <$ 0.05, Wilcoxon signed rank test) after optimal reconstruction of the data, indicating higher identifiability of the subjects among the cohort.
This result held for both the \textit{NoGSR} and \textit{GSR} cases.

\begin{figure}
\centering
\includegraphics[scale =0.9,trim= {3cm 10cm 2cm 11cm}, clip=true]{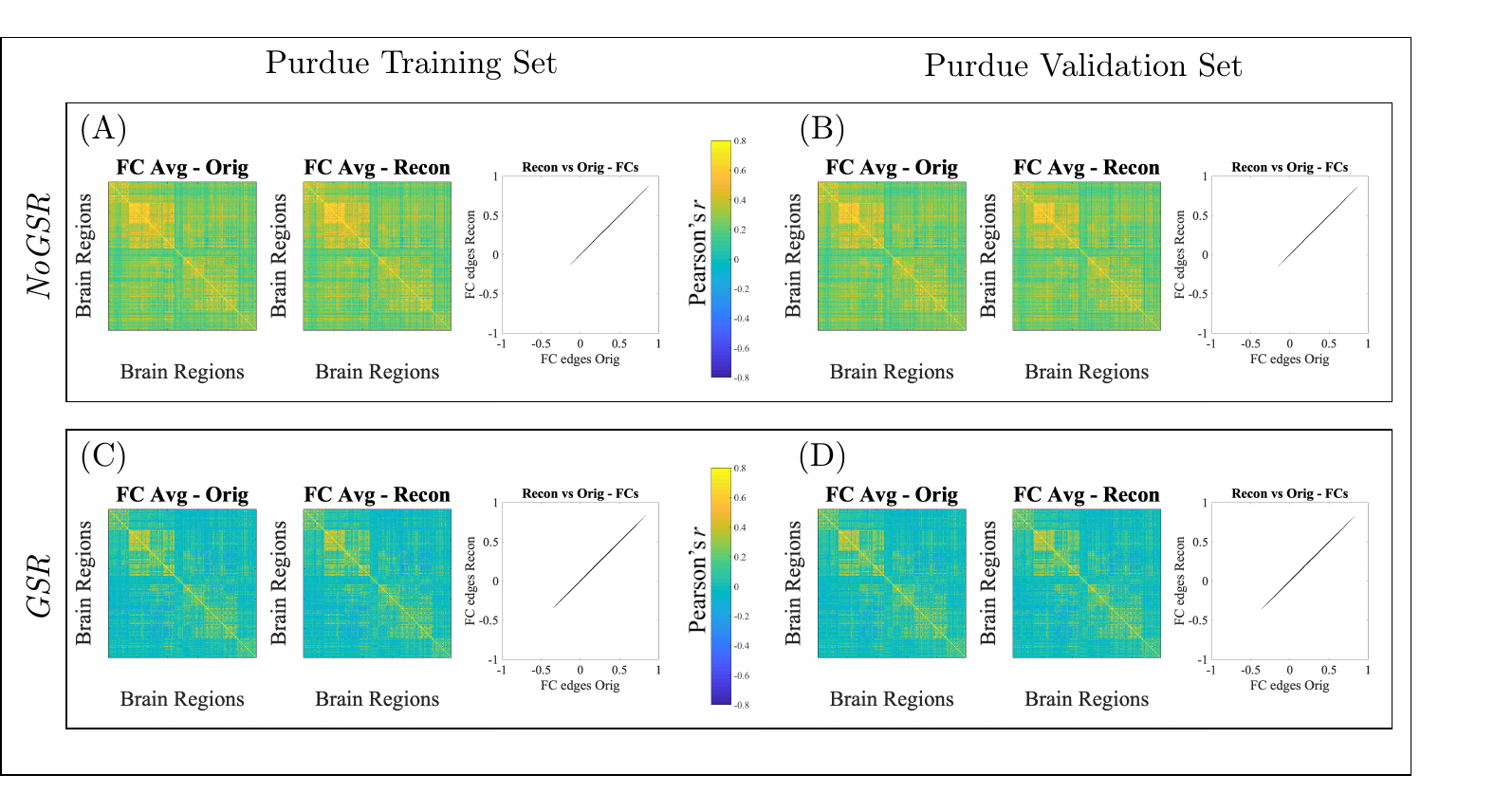}
\caption{Purdue dataset. Evaluation of PCA reconstruction at the optimal number of components (\textbf{m*} = 21) for resting-state functional connectomes (FCs) data without global signal regression (\textit{NoGSR}; (A) and (B)) and  (\textbf{m*} = 22) with global signal regression (\textit{GSR}; (C) and (D)). Left-to-right in each of (A)-(D): the group averaged FC of the original (Orig) data; the group averaged FC of the reconstructed (Recon) data; the scatter plot (for all edges) of the Recon group-averaged FC (y-axis) vs. the Orig group-averaged FC (x-axis). 
Again, exclusion/inclusion of global signal regression did not alter the benefit of the reconstruction to enhance identifiability.}
\label{av_data}
\end{figure}

The group averages of original and optimally reconstructed FCs using \textbf{m*} PCs were computed. Figure \ref{av_data} shows that the group average of original and reconstructed functional connectomes were almost identical, indicating that the optimal PCA reconstruction preserved the main group-level characteristics of the functional connectomes for both \textit{NoGSR} (Figure \ref{av_data} A-B) and \textit{GSR} (Figure \ref{av_data} C-D) datasets.

\begin{figure}
\centering
\includegraphics[scale =0.9,trim= {3cm 6.5cm 2cm 7.5cm}, clip=true]{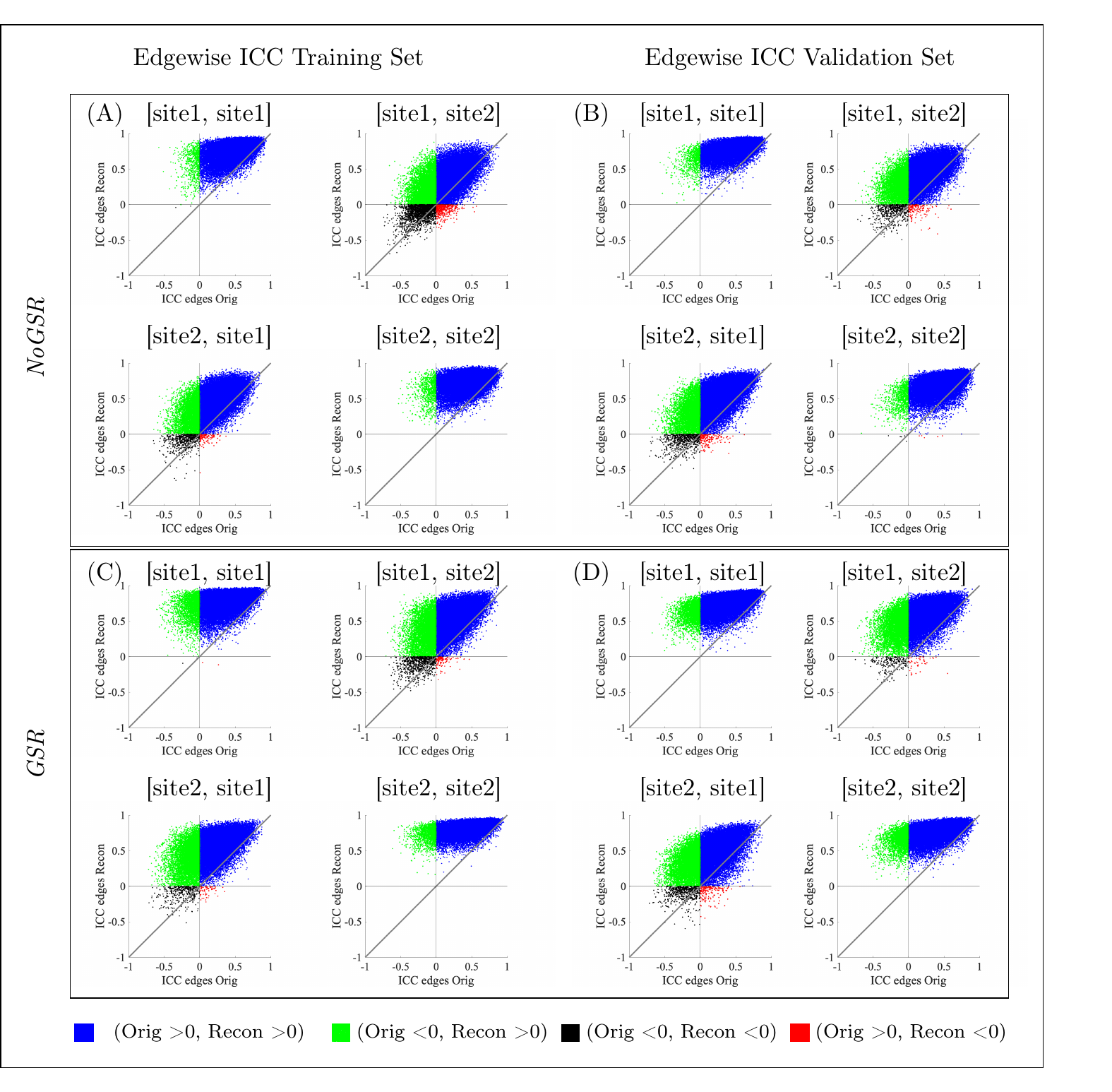}
\caption{Purdue dataset. Scatter plots of averaged (100 iterations) intra-class correlation coefficient (ICC) values, computed over each FC edge, for the reconstructed (Recon) data (y-axis) versus the edgewise ICC for the original (Orig) data (x-axis). Plots are presented for data without global signal regression (\textit{NoGSR}; (A) and (B))  and with global signal regression (\textit{GSR}; (C) and (D)). In each plot, quadrants are colored for clarity of the effect of reconstruction on ICC values. Blue represents positive values in both Orig and Recon; green represents negative Orig and positive Recon; black represents negative values for both Orig and Recon; and red represents positive Orig and negative Recon. 
Note that the vast majority of ICC values have been made more positive by the reconstruction process.}
\label{ICC_scatter}
\end{figure}

\begin{figure}
\centering
\includegraphics[scale =0.9,trim= {3cm 8cm 2cm 9cm}, clip=true]{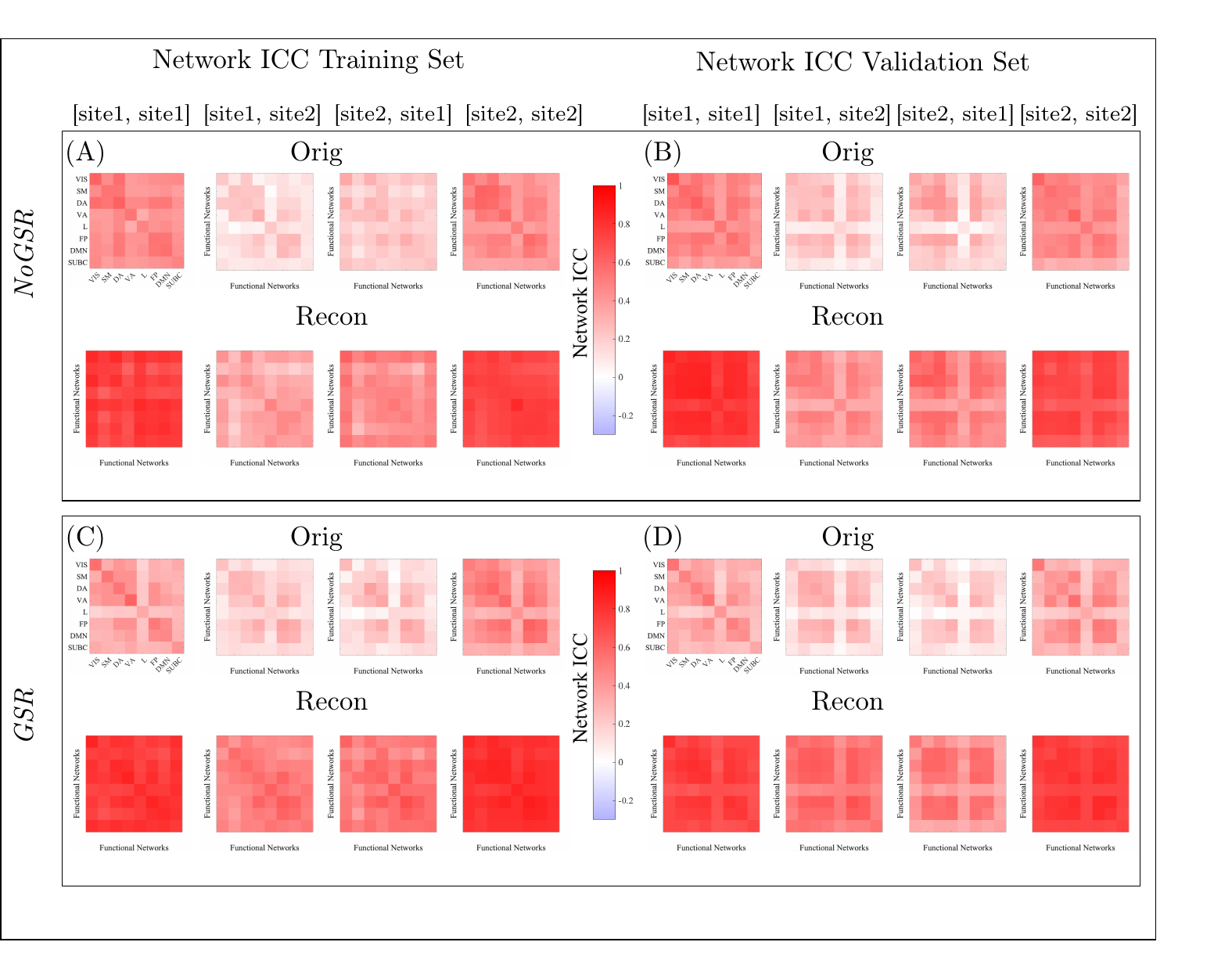}
\caption{Purdue dataset. Intra-class correlation coefficient (ICC) values for each functional network, computed as the average of edgewise ICC over each of Yeo's resting-state functional networks in the original (Orig) and reconstructed (Recon) data for \textit{Training} and \textit{Validation} sets on resting-state functional connectomes without global signal regression (\textit{NoGSR}; (A) and (B)) and with global signal regression (\textit{GSR}; (C) and (D)). Yeo's resting functional networks \cite{Yeo2011TheConnectivity}: Visual (VIS), Somato-Motor (SM), Dorsal Attention (DA), Ventral Attention (VA), Limbic system (L), Fronto-Parietal (FP), Default Mode Network (DMN), and subcortical regions (SUBC).
Once again, no meaningful effect of exclusion/inclusion of global signal regression is observed on the benefit from reconstruction to enhance identifiability.}
\label{ICC_network}
\end{figure}

\begin{figure}[h!]
\centering
\includegraphics[scale =0.9,trim= {3cm 8cm 2cm 9cm}, clip=true]{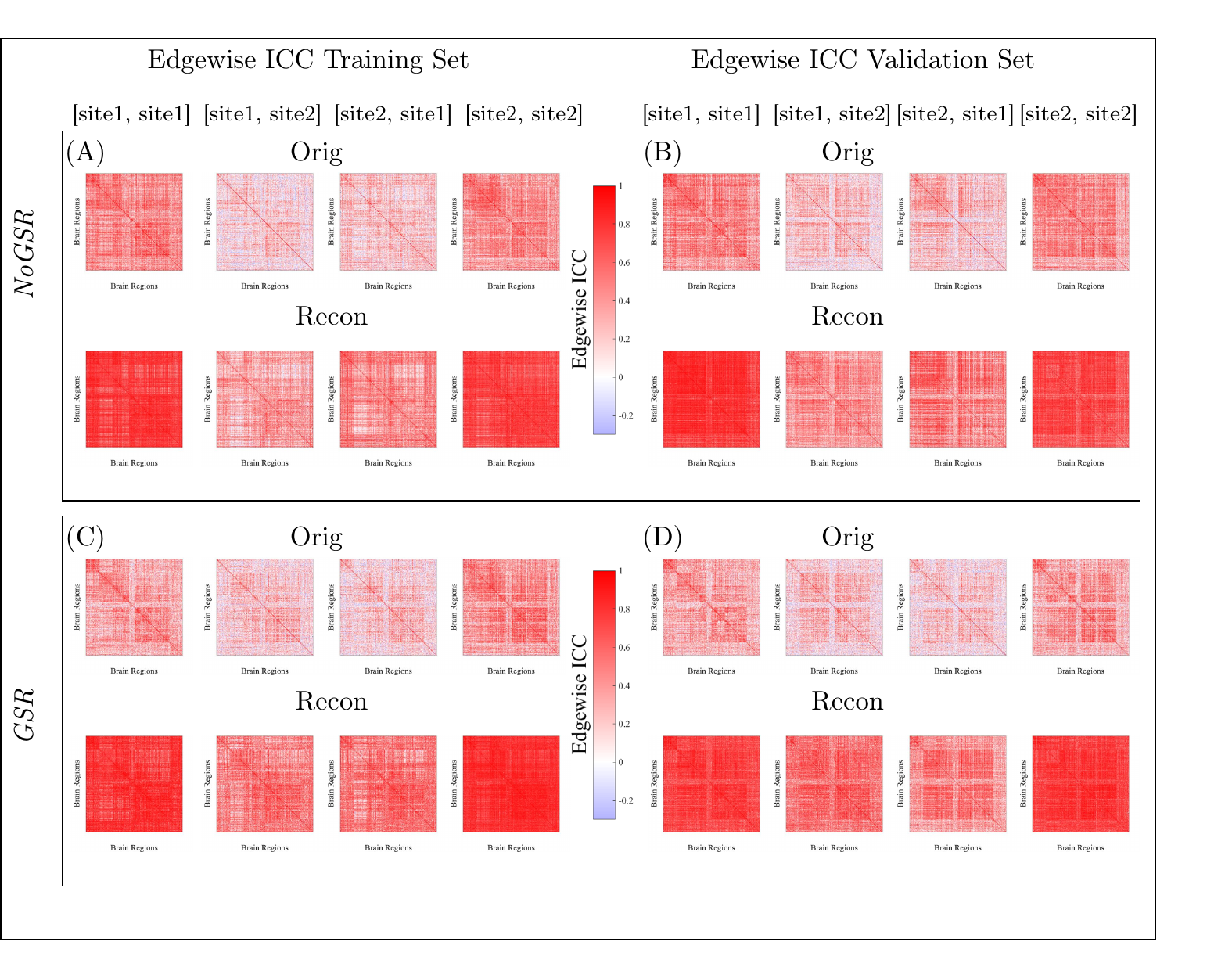}
\caption{Purdue dataset. Averaged (100 iterations) intra-class correlation coefficient (ICC) values, computed for each FC edge, for the original (Orig) and reconstructed (Recon) on the \textit{Training} and \textit{Validation} sets, for resting-state functional connectomes without global signal regression (\textit{NoGSR}; (A) and (B)) and with global signal regression (\textit{GSR}; (C) and (D)). Edges are arranged by Yeo's resting-state functional networks \cite{Yeo2011TheConnectivity}. 
As before, notable benefit is observed for the reconstruction to enhance identifiability, independent of the exclusion/inclusion of global signal regression.}
\label{ICC_edgewise}
\end{figure}

ICC was used to assess the reproducibility of edges in functional connectomes between visits of subjects within and across the two sites. The average ICC value, over 100 iterations obtained from the bootstrap procedure (see Methods for details), from original and optimally reconstructed FCs were compared. ICC for each functional network was computed by averaging over ICC values for all the edges that belonged to a functional network. 
Figures \ref{ICC_scatter},\ref{ICC_edgewise} show the edgewise ICC averaged over 100 iterations for the original and the reconstructed data. The edgewise ICC largely increased after optimal reconstruction for almost all edges (Tables \ref{NoGSRtable}, \ref{GSRtable}) for each [site\textit{i}, site\textit{j}] pair for \textit{NoGSR} (Figures \ref{ICC_scatter},\ref{ICC_edgewise} A-B) and \textit{GSR} (Figures \ref{ICC_scatter},\ref{ICC_edgewise} C-D) datasets. 
Figure \ref{ICC_network} shows that the average ICC for each functional network in the reconstructed data was also higher than in the original data. 

\begin{figure}
\centering
\includegraphics[scale =0.9,trim= {3cm 9.5cm 2cm 10.5cm}, clip=true]{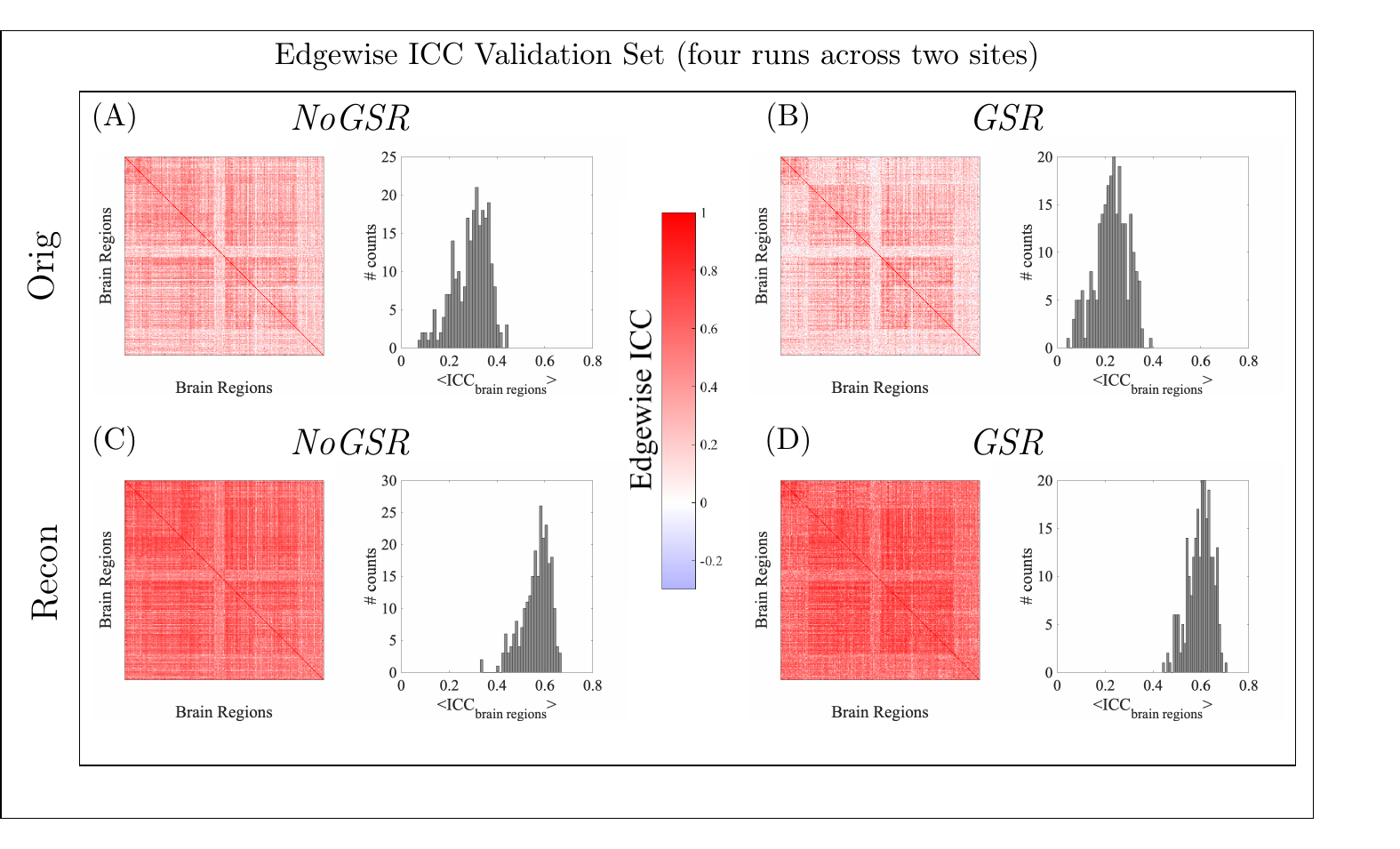}
\caption{Purdue dataset. Averaged (100 iterations; see Methods for bootstrap details) intra-class correlation coefficient (ICC) values, computed for each FC edge from four visits across two sites, for the \textit{Validation} set original (Orig; (A) and (B)) and reconstructed (Recon; (C) and (D)) data without global signal regression (\textit{NoGSR}; (A) and (C)) and with global signal regression (\textit{GSR}; (B) and (D)).
Note that the benefit from reconstruction to enhance identifiability is, again, not dependent on exclusion/inclusion of global signal regression.}
\label{ICC_four_visit}
\end{figure}

\begin{figure}
\centering
\includegraphics[scale =0.9,trim= {3cm 8cm 2cm 9cm}, clip=true]{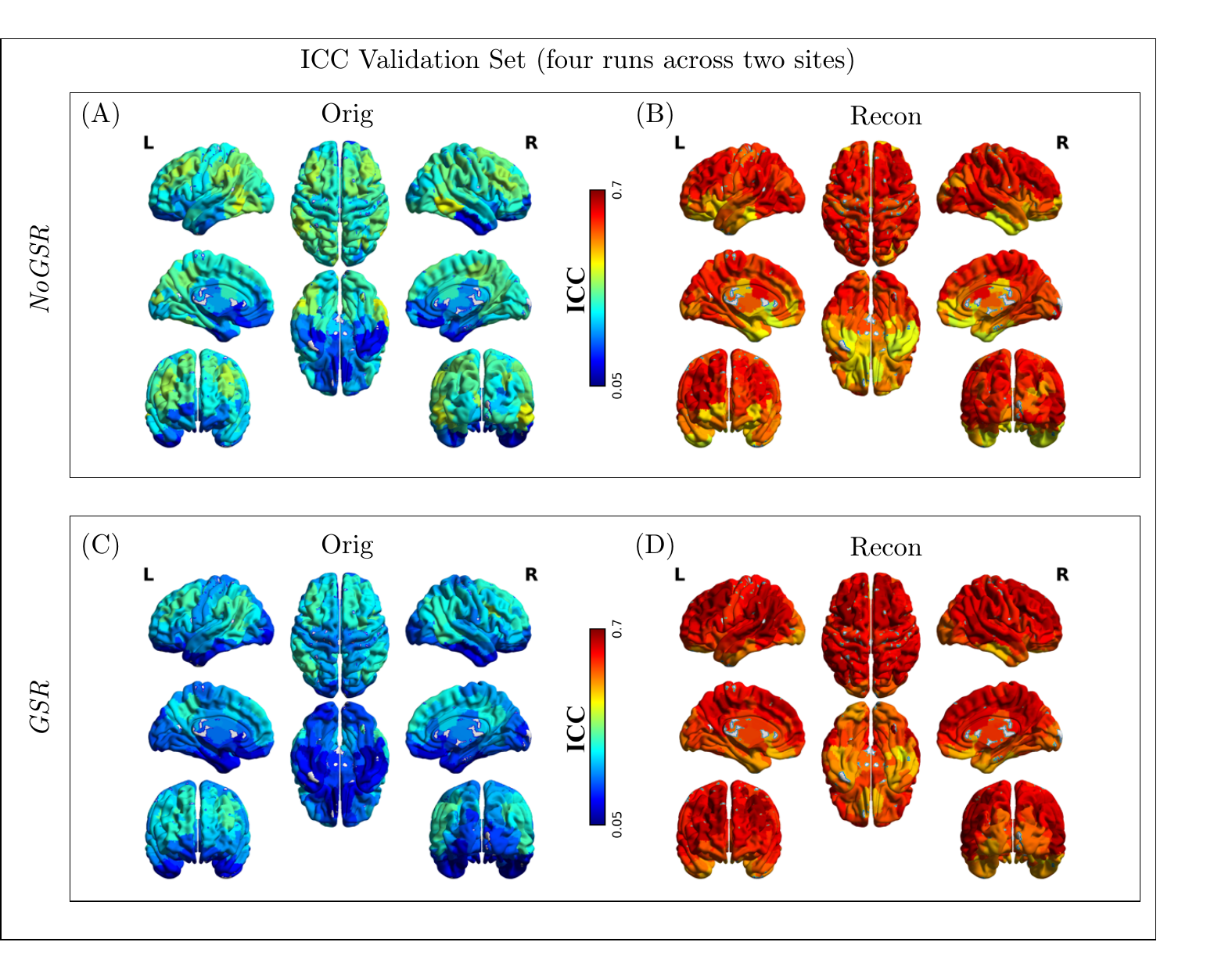}
\caption{Purdue dataset. Brain rendering of intraclass correlation coefficient (ICC), computed from all four visits across the two sites for  the \textit{Validation} set original (Orig; (A) and (C)) and reconstructed (Recon; (B) and (D)) data without global signal regression (\textit{NoGSR}; (A) and (B)) and with global signal regression (\textit{GSR}; (C) and (D)). The strength per brain region---computed as the mean of edgewise ICC values (ICC computed for each FC edge and averaged over 100 iterations; see Methods for Bootstrap procedure)---provides an assessment of overall reproducibility of the functional connections of each brain region.
FC reproducibility was appreciably improved, regardless of exclusion/inclusion of global signal regression.} 
\label{ICC_brain}
\end{figure}

When integrating test-retest FC data from both imaging sites, we measured edgewise ICC, pooling all four visits per subject. 
Figure \ref{ICC_four_visit} shows the edgewise ICC and histograms for average ICC for each brain region (using the atlas from \cite{Shen2013GroupwiseIdentification}) for the original and reconstructed data in the validation set. Figure \ref{ICC_brain} presents a brain rendering overlaid with the averaged edgewise ICC values of each brain region as computed from all four test-retest visits across the two sites using the validation dataset. 
The edgewise ICC and value per brain region for optimally reconstructed data indicated higher reproducibility of the functional connectomes. Both edgewise and average brain region ICC values increased after optimal reconstruction from \textbf{m*} PCs, indicating higher reproducibility and identifiability of the reconstructed functional connectomes as compared to the original ones. 

\begin{figure}[h!]
\centering
\includegraphics[scale =0.8,trim= {3cm 9.5cm 2cm 10.5cm}, clip=true]{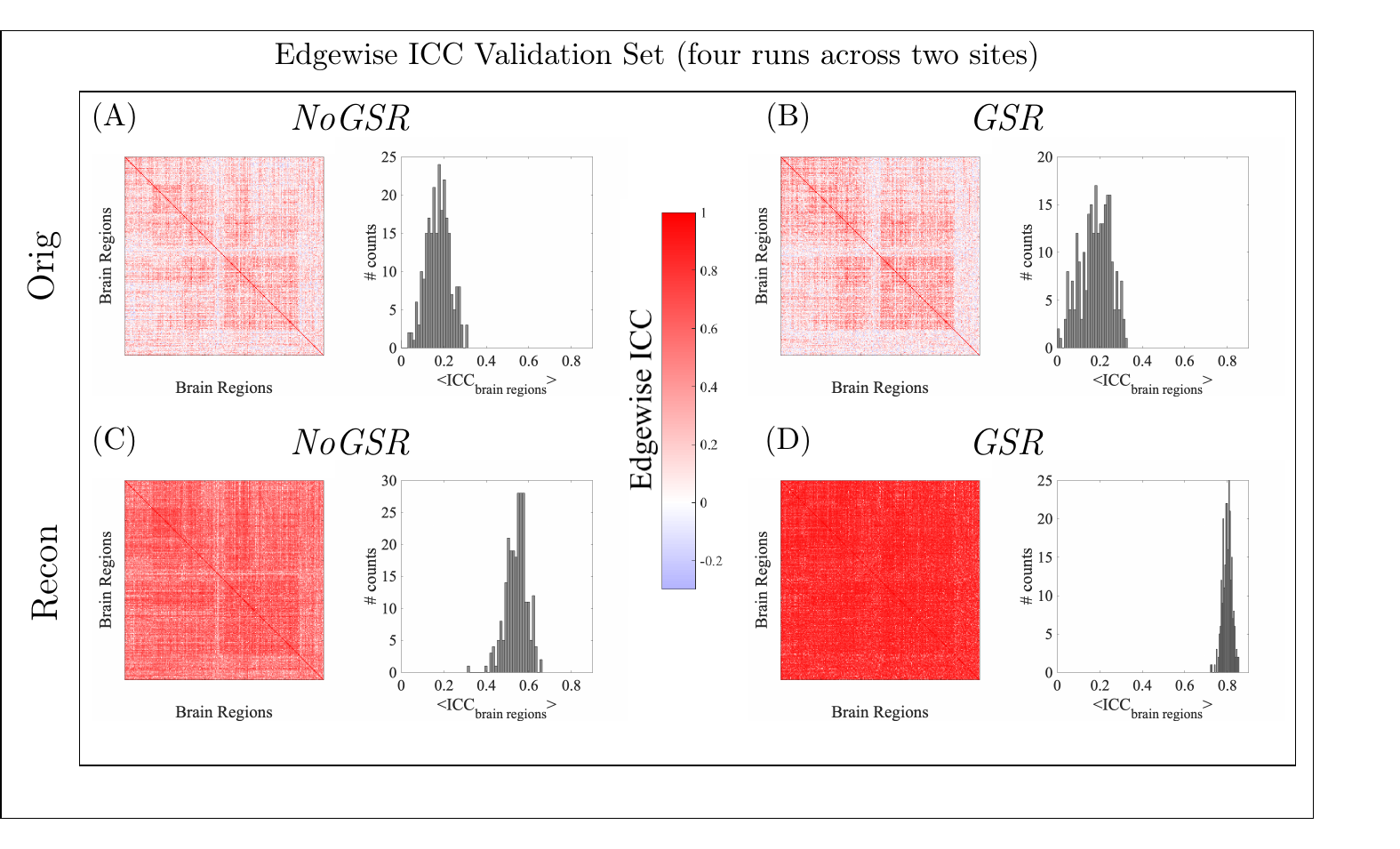}
\caption{Yale dataset. Averaged (100 iterations; see Methods for bootstrap details) intra-class correlation coefficient (ICC) values, computed for each FC edge from four visits across two sites, for the \textit{Validation} set original (Orig; (A) and (B)) and reconstructed (Recon; (C) and (D)) data without global signal regression (\textit{NoGSR}; (A) and (C)) and with global signal regression (\textit{GSR}; (B) and (D)).
Note that the benefit from reconstruction to enhance identifiability is, again, not dependent on exclusion/inclusion of global signal regression.}
\label{ICC_four_visit_yale}
\end{figure}

\begin{figure}[h!]
\centering
\includegraphics[scale =0.8,trim= {3cm 8cm 2cm 9cm}, clip=true]{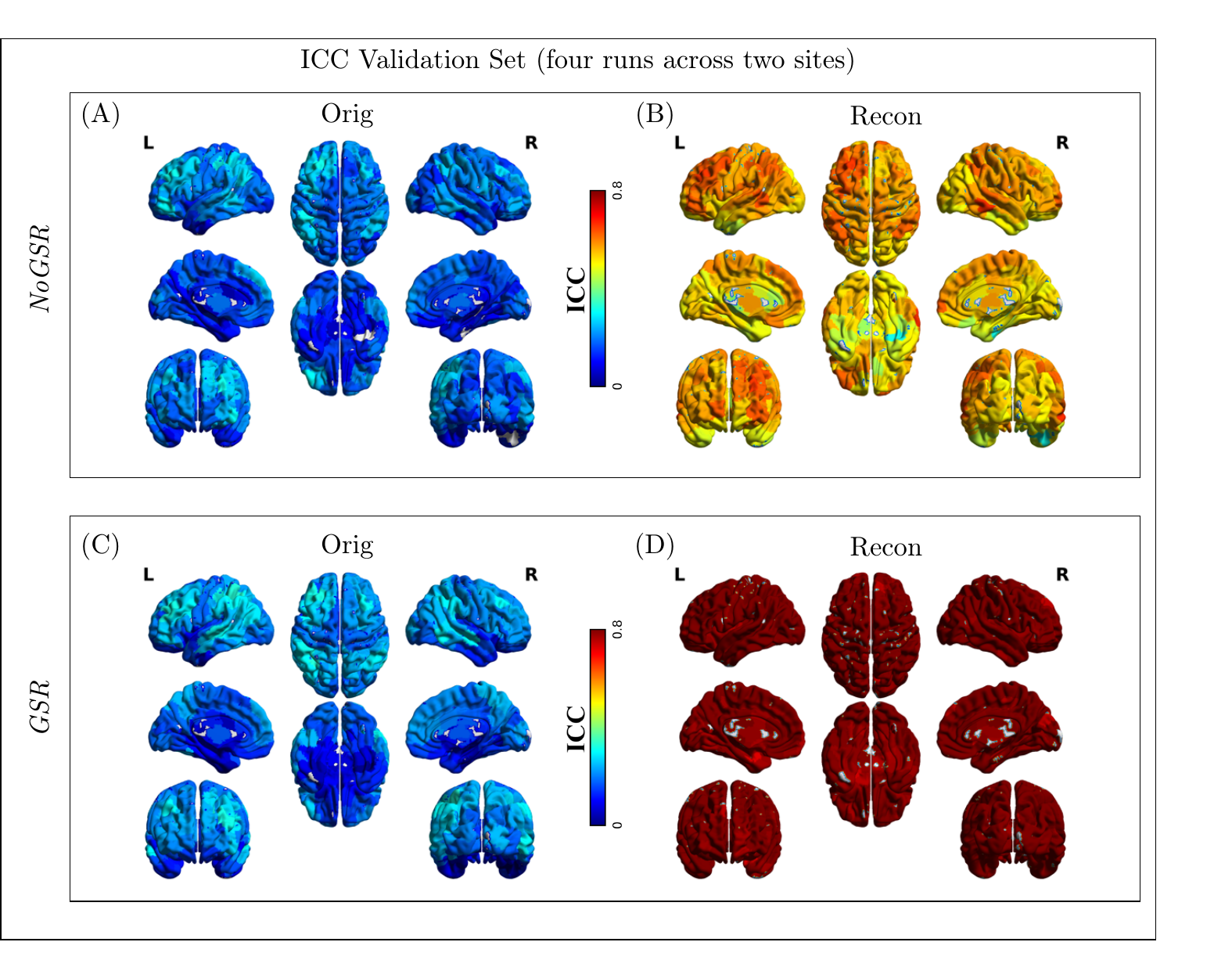}
\caption{Yale dataset. Brain rendering of intraclass correlation coefficient (ICC), computed from all four visits across the two sites for  the \textit{Validation} set original (Orig; (A) and (C)) and reconstructed (Recon; (B) and (D)) data without global signal regression (\textit{NoGSR}; (A) and (B)) and with global signal regression (\textit{GSR}; (C) and (D)). The strength per brain region---computed as the mean of edgewise ICC values (ICC computed for each FC edge and averaged over 100 iterations; see Methods for Bootstrap procedure)---provides an assessment of overall reproducibility of the functional connections of each brain region.
FC reproducibility was appreciably improved, regardless of exclusion/inclusion of global signal regression.} 
\label{ICC_brain_yale}
\end{figure}

Notably, all these findings were replicated in the Yale dataset. The results obtained for the Yale dataset are shown in Supplementary material (see Figures \ref{Idiff_yale}, \ref{Ident_matrices_yale}, \ref{ICC_four_visit_yale} and \ref{ICC_brain_yale} and Table \ref{I_diff_ij_yale}). Specifically, Figure \ref{Idiff_yale} shows $<<I_{diff}>>$ as a function of the number of PCs (\textbf{m}) and it peaks at \textbf{m*} = 12 for both \textit{NoGSR} and \textit{GSR}. The identifiability matrices obtained from reconstructed FCs using these \textbf{m*} PCs as compared to the original ones are shown in Figure \ref{Ident_matrices_yale}. Figures \ref{ICC_four_visit_yale}, \ref{ICC_brain_yale} depicts the edgewise ICC results when pooling all four visits together.

\begin{figure}
\centering
\includegraphics[scale =0.9,trim= {3cm 7.5cm 2cm 8.5cm}, clip=true]{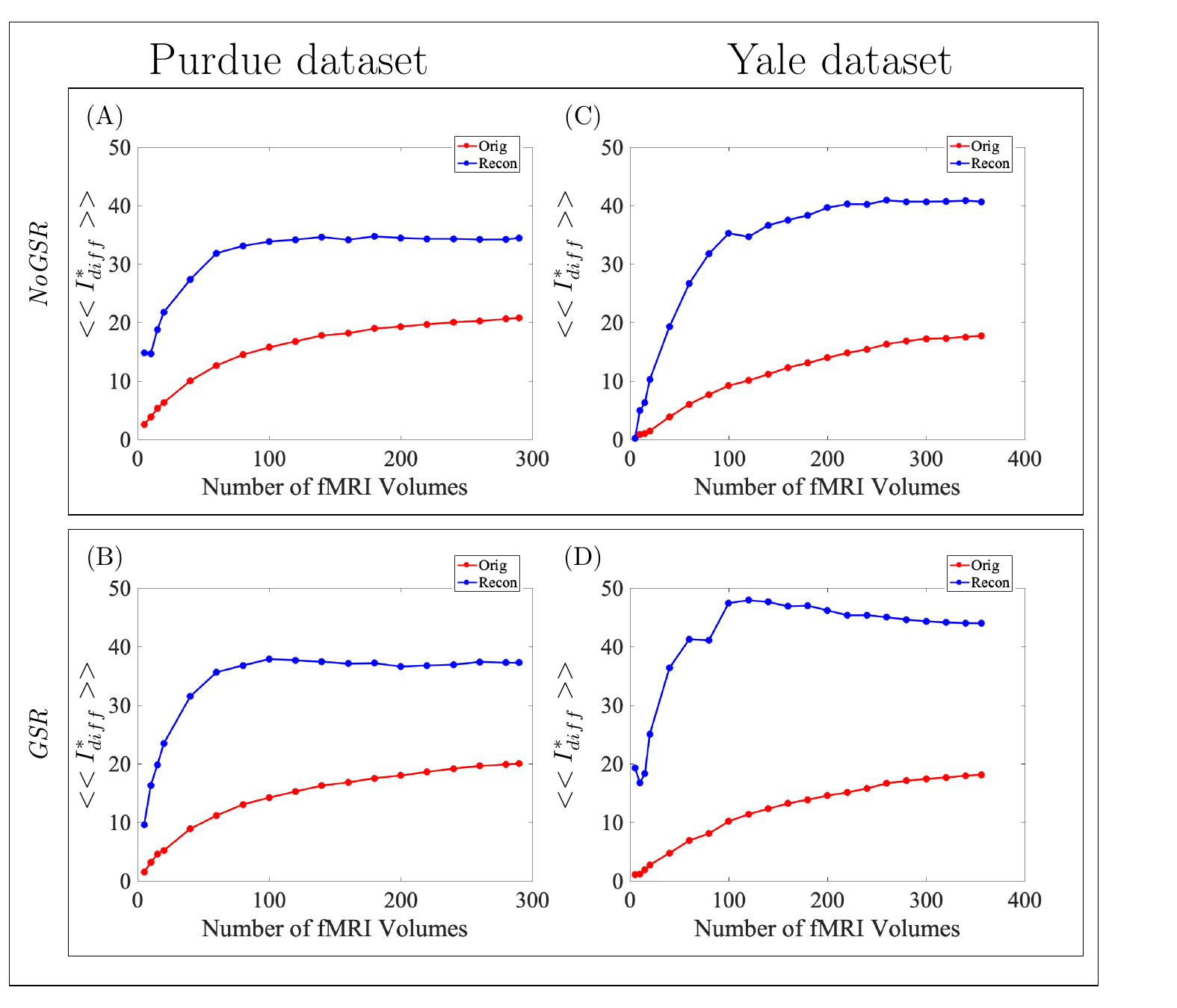}
\caption{Optimal multi-site differential identifiability $(<<I_{diff}^{*}>>*100)$ as a function of the number of fMRI volumes used for reconstruction for resting-state Purdue and Yale datasets without global signal regression (\textit{NoGSR}; (A) and (C)) and with global signal regression (\textit{GSR}; (B) and (D)). 
It may be observed that the benefit of reconstruction on differential identifiability was not dependent on the exclusion/inclusion of global signal regression.}
\label{fMRI_Idiff}
\end{figure}

The effect of number of fMRI volumes on multi-site differential identifiability was assessed. To that end, processed BOLD time-series were shortened (by dropping fMRI volumes) to mimic different scan lengths. For each scan length evaluated, FCs were estimated, decomposed and subsequently reconstructed based on \textbf{m*} PCs.  Optimal multi-site differential identifiability ($<<I_{diff}^{*}>>$) was computed from optimally reconstructed FCs and compared to that obtained from original FCs. Figure \ref{fMRI_Idiff} shows that the method presented in this study improved $<<I_{diff}^{*}>>$ for both Purdue and Yale datasets for \textit{NoGSR} and \textit{GSR} for all scan lengths evaluated.

\begin{figure}
\centering
\includegraphics[scale =0.9,trim= {3cm 7.5cm 2cm 8.5cm}, clip=true]{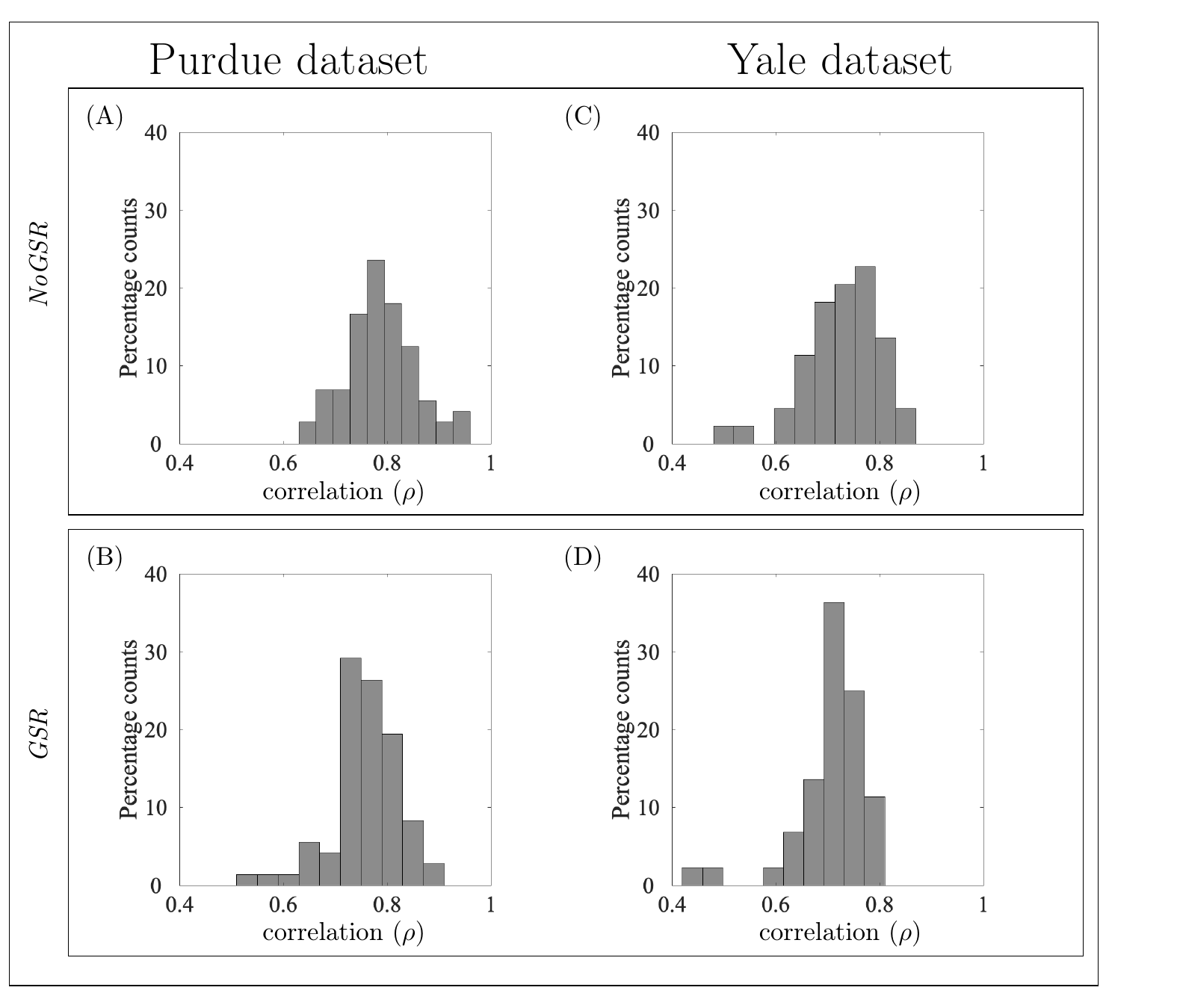}
\caption{\small{ Histograms of similarity between optimally reconstructed FCs (complete dataset for PCA framework) and with leave-one-out (LOO) reconstructed FCs. (A) Purdue without global signal regression (\textit{NoGSR}), (B) Purdue with global signal regression (\textit{GSR}), (C) Yale \textit{NoGSR} and (D) Yale \textit{GSR}}.
}
\label{rho}
\end{figure}

In order to assess the generalizability of the optimal orthogonal basis for each dataset, a leave-one-out experiment was performed. Briefly, each subject's FCs were reconstructed using the \textbf{m*} PCs when all the sessions of that subject were excluded from the PCA framework. For each dataset (Purdue and Yale), the optimally reconstructed FCs of each subject were compared to the leave-one-out reconstructed FCs. Histograms of the correlations of optimally reconstructed FCs from training vs leave-one out for all subjects are shown for Purdue and Yale datasets in the Figure \ref{rho}. Median values were $0.79$ for Purdue \textit{NoGSR}, $0.77$ for Purdue \textit{GSR}, $0.74$ for Yale \textit{NoGSR} and $0.73$ for Yale \textit{GSR}.



\begin{table}[]

\caption{Purdue dataset. Maximum percentage differential identifiability ($<I_{diff}^{ij*}>*100$), number of principal components for each [site\textit{i}, site\textit{j}] pair ($m^{ij*}$), explained variance ($R^{2}$), mean ($\mu$) and standard deviation ($\sigma$) of edgewise ICC values for Original (Orig) and optimally reconstructed (Recon) for Training datasets without global signal regression (\textit{NoGSR}) and with global signal regression (\textit{GSR}).} 
\begin{tabular}{|c|c|c|c|c|c|c|c|c|}
\hline
                       & [site\textit{i}, site\textit{j}] & $<I_{diff}^{ij*}>$ & $m^{ij*}$  & $R^{2}$ &ICC $\mu_{Orig}$ &ICC $\sigma_{Orig}$ & ICC $\mu_{Recon}$ & ICC $\sigma_{Recon}$\\ \hline
\multirow{4}{*}{\textit{NoGSR}} & [site1, site1]        & 45.1     & 29  & 0.82 & 0.46& 0.20 & 0.75 & 0.12\\ \cline{2-9} 
                       & [site1, site2]        & 31.0     & 20  &  0.74 & 0.14 & 0.24 & 0.34 & 0.22\\ \cline{2-9} 
                       & [site2, site1]        & 31.5     & 20  & 0.74 & 0.20 & 0.23 & 0.44 & 0.22\\ \cline{2-9} 
                       & [site2, site2]        & 45.7     & 31 & 0.84 & 0.44 & 0.20& 0.72 & 0.11\\ \hline
\multirow{4}{*}{\textit{GSR}}   & [site1, site1]        & 43.6     & 35  & 0.84 & 0.35& 0.24 & 0.77 & 0.13\\ \cline{2-9} 
                       & [site1, site2]        & 30.8     & 21  & 0.73 & 0.18 & 0.25 & 0.49 & 0.22 \\ \cline{2-9} 
                       & [site2, site1]        & 32.8     & 21  & 0.73 & 0.17& 0.25 & 0.53 & 0.20\\ \cline{2-9} 
                       & [site2, site2]        & 45.5     & 35  & 0.84 & 0.42 & 0.23 & 0.81 & 0.08\\ \hline
\end{tabular}
\label{I_diff_ij}
\end{table}

\begin{table}[]

\caption{Purdue dataset. Percentage of positive and negative edgewise intra-class correlation coefficient (ICC) values (computed for each FC edge and averaged over 100 iterations; see Methods for Bootstrap procedure) of original (Orig) data that were converted to positive or negative edgewise ICC in reconstructed (Recon) data for resting-state functional connectomes without global signal regression (\textit{NoGSR}).}
\begin{tabular}{|c|c|c|c|c|c|c|}
\hline
[site\textit{i}, site\textit{j}]                               & \multicolumn{3}{c|}{Purdue Training Set}      & \multicolumn{3}{c|}{Purdue Validation Set}     \\ \hline
\multirow{4}{*}{{[}site1, site1{]}} &           & \multicolumn{2}{c|}{Recon} &            & \multicolumn{2}{c|}{Recon} \\ \cline{2-7} 
                                    & Orig      & Negative     & Positive    & Orig       & Negative     & Positive    \\ \cline{2-7} 
                                    & Negative  & 0.28         & 99.72           & Negative   & 0.00     & 100.00           \\ \cline{2-7} 
                                    & Positive  & 0.0         & 100.00           & Positive   & 0.00     & 100.0           \\ \hline
\multirow{4}{*}{{[}site1, site2{]}} &           & \multicolumn{2}{c|}{Recon} &            & \multicolumn{2}{c|}{Recon} \\ \cline{2-7} 
                                    & Orig      & Negative     & Positive    & Orig       & Negative     & Positive    \\ \cline{2-7} 
                                    & Negative  & 20.56        & 79.44       & Negative   & 7.72         & 92.28           \\ \cline{2-7} 
                                    & Positive  & 1.85         & 98.15       & Positive   & 0.65         & 99.34           \\ \hline
\multirow{4}{*}{{[}site2, site1{]}} &           & \multicolumn{2}{c|}{Recon} &            & \multicolumn{2}{c|}{Recon} \\ \cline{2-7} 
                                    & Orig      & Negative     & Positive    & Orig       & Negative     & Positive    \\ \cline{2-7} 
                                    & Negative  & 9.75        & 90.25           & Negative   & 10.50            & 89.50           \\ \cline{2-7} 
                                    & Positive  & 0.47         & 99.53           & Positive   & 0.72            & 99.28           \\ \hline
\multirow{4}{*}{{[}site2, site2{]}} &           & \multicolumn{2}{c|}{Recon} &            & \multicolumn{2}{c|}{Recon} \\ \cline{2-7} 
                                    & Orig      & Negative     & Positive    & Orig       & Negative     & Positive    \\ \cline{2-7} 
                                    & Negative  & 0.00            & 100.00           & Negative   & 0.61            & 99.39           \\ \cline{2-7} 
                                    & Positive  & 0.00            & 100.00           & Positive   & 0.02            & 99.98           \\ \hline
\end{tabular}
\label{NoGSRtable}
\end{table}
\begin{table}[]

\caption{Purdue dataset. Percentage of positive and negative edgewise intra-class correlation coefficient (ICC) values (computed for each FC edge and averaged over 100 iterations; see Methods for Bootstrap procedure) of original (Orig) data that were converted to positive or negative edgewise ICC in reconstructed (Recon) data for resting-state functional connectomes with global signal regression (\textit{GSR}).}
\begin{tabular}{|c|c|c|c|c|c|c|}
\hline
[site\textit{i}, site\textit{j}]                               & \multicolumn{3}{c|}{Purdue Training Set}      & \multicolumn{3}{c|}{Purdue Validation Set}     \\ \hline
\multirow{4}{*}{{[}site1, site1{]}} &           & \multicolumn{2}{c|}{Recon} &            & \multicolumn{2}{c|}{Recon} \\ \cline{2-7} 
                                    & Orig      & Negative     & Positive    & Orig       & Negative     & Positive    \\ \cline{2-7} 
                                    & Negative  & 0.08         & 99.92       & Negative   & 0.00         & 100.00           \\ \cline{2-7} 
                                    & Positive  & 0.01         & 99.99       & Positive   & 0.00         & 100.00           \\ \hline
\multirow{4}{*}{{[}site1, site2{]}} &           & \multicolumn{2}{c|}{Recon} &            & \multicolumn{2}{c|}{Recon} \\ \cline{2-7} 
                                    & Orig      & Negative     & Positive    & Orig       & Negative     & Positive    \\ \cline{2-7} 
                                    & Negative  & 10.61         & 89.39           & Negative   & 2.78     & 97.22           \\ \cline{2-7} 
                                    & Positive  & 0.46         & 99.54           & Positive   & 0.18     & 99.82           \\ \hline
\multirow{4}{*}{{[}site2, site1{]}} &           & \multicolumn{2}{c|}{Recon} &            & \multicolumn{2}{c|}{Recon} \\ \cline{2-7} 
                                    & Orig      & Negative     & Positive    & Orig       & Negative     & Positive    \\ \cline{2-7} 
                                    & Negative  & 4.80         & 95.20           & Negative   & 6.34     & 93.66           \\ \cline{2-7} 
                                    & Positive  & 0.27         & 99.73           & Positive   & 0.80     & 99.20           \\ \hline
\multirow{4}{*}{{[}site2, site2{]}} &           & \multicolumn{2}{c|}{Recon} &            & \multicolumn{2}{c|}{Recon} \\ \cline{2-7} 
                                    & Orig      & Negative     & Positive    & Orig       & Negative     & Positive    \\ \cline{2-7} 
                                    & Negative  & 0.00            & 100.00           & Negative   & 0.00       & 100.00           \\ \cline{2-7} 
                                    & Positive  & 0.00            & 100.00           & Positive   & 0.00          & 100.00           \\ \hline
\end{tabular}
\label{GSRtable}
\end{table}
\section{Discussion}
\label{discussion}

Recently the concepts of brain fingerprinting and identifiability \cite{Mars2018ConnectivitySpaces} have been investigated based on repeated measures of individual whole-brain estimates of resting-state functional connectivity \cite{Mira-Dominguez2014Connectotyping:Connectome,Finn2015FunctionalConnectivity} and between fMRI tasks \cite{Finn2017CanConnectivity,Greene2018Task-inducedTraits,Yoo2018Connectome-basedDatasets}. More recently, Amico et al. \cite{Amico2018TheConnectomes} introduced the concept of an identifiability matrix to assess the fingerprinting of a dataset through a functional denominated identifiability score (see Methods). Further they introduced a data-driven method to uncover identifiability in whole-brain functional connectomes (FCs) based on principal component decomposition and subsequent reconstruction. Here, we extended this framework for multi-site repeated measurements experiments and show how high identifiability on an inter-scanner basis is achievable at the whole-brain level, as well as at the pairwise level for functional edges. This approach to uncover identifiability was equally effective for rs-fMRI data processed with and without global signal regression. Results indicate that the individual fingerprints obtained from optimally reconstructed FCs were robust, and improved identifiability among the cohort. Further, the method improved the reproducibility of the functional connectivity profiles across visits, both on an edgewise and functional network basis. We discuss below all the results related to the Purdue dataset.

Multi-site differential identifiability $<<I_{diff}>>$ was used as a quality function to maximize the fingerprinting of individual subjects within a cohort by exploring connectivity subspaces over a range of M principal components. The identifiability of a connectivity profile of a subject relies on the fact that individual subjects are expected to be most similar to themselves across visits or scanning sessions, relative to others. We used a continuous identifiability score as defined by \cite{Amico2018TheConnectomes} for individual fingerprinting of subjects in test-retest sessions for two sites. The continuous identifiability score, $<I_{diff}^{ij}>$, quantified the difference between average within-subject similarity and average between-subject similarity for a single [site\textit{i}, site\textit{j}] visit pair. $<<I_{diff}>>$ quantified the overall fingerprinting of the population across all test-retest visits. $<<I_{diff}>>$ was then maximized over subsets of M PCs to find the \textbf{m*} PCs that maximized differential identifiability and provided the optimal orthogonal basis to reconstruct the FCs. For both the \textit{NoGSR} and \textit{GSR} datasets, $<<I_{diff}>>$ and $<I_{diff}^{ij}>$ (Figure \ref{Idiff}) showed a significant improvement over the identifiability score computed from the original FCs. The higher value of average differential identifiability indicates stronger overall individual fingerprinting of the population. 

When assessing $<I_{diff}^{ij}>$ and $m^{ij*}$ (see Table \ref{I_diff_ij}), it can be seen that there are differences in the proportion of the dimensionality of the data that are kept for maximizing identifiability. In particular, visit pairs including different sites (i.e., [site1, site2] and [site2, site1]) had $m^{ij*}$ values very close to the number of subjects (i.e., 16) whereas visit pairs including just one site (i.e., [site1, site1] and [site2, site2] were able to keep a larger number of components, indeed approximately the number visits within an imaging site (i.e., twice the number of subjects). These results emphasize how important it is to formalize a data-driven framework for reconstruction of FCs that is not based on fixing certain number of components or ultimately a percentage of variance, since identifiability might peak at very different configurations depending on multiple factors, including number of subjects, number of imaging sites, baseline similarity between test-retest on the sites, etc.

Identifiability matrices (\textbf{I}) computed from optimally reconstructed data outperformed those computed from the original data. Identifiability matrices consisted of Pearson’s correlation coefficient between FCs corresponding to subjects' test and retest visits, within and across the two sites. The main diagonal of each of the four blocks ($\textbf{I}^{ij}$) consisted of correlations between visits of the same subject within and across the two sites. These self correlations had higher values in \textbf{I} obtained from optimally reconstructed data as compared to the ones obtained from original data (Figure \ref{Ident_matrices}). One of the noteworthy facts about Figure \ref{Ident_matrices} is the substantial increase in self correlations for the challenging problem of test-retest visits between the two sites. This indicates stronger individual fingerprints of subjects after optimal reconstruction of the FCs, not only in repeated visits within the same site, but also among visits across two sites.   

Statistically higher values for the distributions of $\textbf{I}_{diff}^{ij}$ for all test-retest [site\textit{i}, site\textit{j}] pairs of the reconstructed data as compared to the original data illustrated stronger fingerprinting of the population. $\textbf{I}_{diff}^{ij}$ quantified the differential identifiability on a subject level for the test-retest [site\textit{i}, site\textit{j}] pairs. Higher $\textbf{I}_{diff}^{ij}$ values indicated improved identifiability of subjects. Differential identifiability increased for all subjects within the same site visits, and also between the two sites after optimal reconstruction of the FCs. No difference was found in the reconstructed data between distributions for \textit{NoGSR} and \textit{GSR} $\textbf{I}_{diff}^{ij}$, suggesting that both approaches benefit from this framework, both within and between-sites in a similar way. 

The group averages of the original and optimally reconstructed FCs were almost identical, indicating that the main group level features of the functional connectivity profiles were preserved by the optimal reconstruction. The \textbf{m*} PCs that maximized the $<<I_{diff}>>$ obtained from the training data, were used as an optimal orthogonal basis to reconstruct the functional connectivity profiles for subjects' test and retest sessions in both the training and validation sets. In general PCA is used to transform a set of correlated variables into a set of linearly uncorrelated variables, namely the principal components. The principal components are arranged in descending order of their explained variance and provide a new basis to represent the data. Keeping the subset of the first \textbf{m*} PCs helps to provides a simpler representation of the data through dimensionality reduction while still retaining critical information. Here we rely on the fact, as pointed out by \cite{Amico2018TheConnectomes},
that the highest variance principal components carry cohort-level functional connectivity information, while lower variance PCs carry finer subject-level functional connectivity information, and the lowest variance PCs carry information regarding noise and artifacts. By using the set of first \textbf{m*} PCs, which maximized averaged differential identifiability, for reconstruction provided a denoised version of the original FCs by keeping the cohort and finer subject level functional connectivity information.  

To assess identifiability at a finer grain perspective, we considered the pairwise intraclass correlation coefficient (ICC) at the level of functional edges. Optimally reconstructed FCs systematically showed increased ICC values as compared to original FCs. At the meso-scale of looking at resting-state networks and their interactions, analogous ICC increases were found. In this study the groups were the test-retest visits within a site or between the two sites, whereas the measurements were the values of each functional connectivity edge from all subjects. The reconstructed FCs represent a denoised version of the original data, having lower variance between measurements on different groups. ICC values in Figures \ref{ICC_scatter}-\ref{ICC_brain} and Tables \ref{NoGSRtable}-\ref{GSRtable} indicated higher reproducibility of the functional connectivity profiles after optimal reconstruction. The reproducibility of the edges also helped to distinguish subjects and augment identifiability. In other words, higher ICC values led to higher identifiability of the functional connectivity edges. There was a notable increase in ICC values of the reconstructed data for the challenging problem of between-site test-retest visit pairs. Thus, the significant increase in ICC values for the reconstructed data denoted higher identifiability in all test-retest visit pairs for both \textit{NoGSR} and \textit{GSR} datasets after optimal reconstruction with \textbf{m*} PCs.

Notably, these findings were replicated in a second independent dataset, here denominated the Yale dataset (see Figures \ref{Idiff_yale}, \ref{Ident_matrices_yale}, \ref{ICC_four_visit_yale} and \ref{ICC_brain_yale} and Table \ref{I_diff_ij_yale}). Indeed, $<<I_{diff}>>$ reached even higher values when compared to the Purdue dataset. It is possible that this is because of the identically configured imaging sites as well as because of the shorter TR. Both characteristics might facilitate higher identifiability scores.

When assessing the effect of scan lengths on multi-site differential identifiability, we noticed that the optimally reconstructed FCs systematically provided higher levels of $<<I_{diff}^{*}>>$ for both datasets, with and without GSR, and for all scan lengths evaluated (see Figure \ref{fMRI_Idiff}). These multi-site results are in the line of those observed in \cite{Amico2018TheConnectomes} for single-site evaluations and emphasize not only the importance of maximizing identifiability but also the generalizability of this extended framework to different imaging sites, TRs as well as scan lengths or number of fMRI volumes acquired.

The modest cohort-sizes of the two datasets assessed in this paper (Purdue $N=18$, Yale $N=11$) limited our ability to further explore the universality of the sets of orthogonal basis obtained at the optimal reconstruction levels. A leave-one-out procedure (Figure \ref{rho}) showed that both sets of orthogonal basis had limited generalizability, with Purdue (the largest cohort) having higher generalizability than Yale (the smallest one). These results suggest that cohort-size could be critical for this venue. Further work with large inter-scanner datasets should uncover the plausibility of obtaining truly generalizable sets of orthogonal components for single- and multi-scanner fMRI datasets that would optimally reconstruct unseen subjects while preserving their connectivity fingerprints \cite{Sripada2019BasicConnectomes}.

This work has several limitations. A limitation of the method is that this data-driven method requires the availability of test-retest sessions on all subjects and each site, which is usually not available in cross-sectional clinical studies. A limitation of the study is the modest sample size and small number of available sites. However, several multi-site fMRI studies were performed with less than 10 subjects \cite{Casey1998ReproductibilityTask,Friedman2008Test-retestStudy,Zou2005ReproducibilityNetwork,Noble2017MultisiteConnectivity,Deprez2018Multi-centerMeasures}. A larger multi-site study involving more subjects and sites will help to generalize the results. Further, better acquisition parameters for rs-fMRI may improve the results of the study.

This study expanded on the emerging field of fingerprinting in resting-state functional connectomes (FCs), by opening it to a less controlled scenario wherein repeated measurements are obtained from different imaging sites. To do so, it extended a recently proposed method to assess and ultimately improve identifiability in multi-site studies. Future studies could use this method to examine the reproducibility of fingerprinting in resting-state functional networks and structural connectivity across more than two sites. Another avenue of exploration would be to investigate the reliability of graph theory measures (e.g., clustering coefficient, characteristic path length, modularity, etc.) in the denoised FCs. Further use of this extended PCA methodology could be used to denoise T$_{1}$ and T$_{2}$ structural images at the voxel level by reconstructing test-retest MNI registered volumes at the optimal level of differential identifiability. Another important investigation would be to test the method presented in this study on scanners from different vendors, allowing combination of data for larger multi-site studies. 

Multi-site fMRI studies have great appeal as a means of generating larger datasets, but the site-dependent variability can mask the advantages of such studies. Individual fingerprinting is a critical and emerging field in resting-state functional connectivity. Here we evaluated fingerprinting of the subjects in test-retest visit pairs, within and across two sites. We presented a framework based on principal component analysis to denoise the FCs and improve identifiability. We used principal components that maximized differential identifiability on the \textit{training} data as an orthogonal basis to reconstruct subjects' individual FCs for \textit{training} and \textit{validation} datasets. These optimally reconstructed FCs resulted in substantial improvement in individual fingerprinting within same-site visit pairs and also for the challenging problem of between-site visits, relative to the original data. Optimally reconstructed FCs systematically showed a notable increase in ICC values as compared to the original FCs, at the levels of functional edges, resting-state networks, and network interactions. Results showed that it is possible to use the data-driven method presented in this study to improve identifiability in the functional connectivity domain for multi-site studies. This would pave the way to pool subjects recruited at different sites, allowing for better assessments of brain structure and function in the healthy and diseased brain.


\section{Acknowledgments}
JG  acknowledges  financial  support  from  NIH  R01EB022574, NIH R01MH108467, Indiana Alcohol Research Center P60AA07611, and the Indiana Clinical and Translational Sciences Institute (Grant Number UL1TR001108) from the National Institutes of Health, National Center for Advancing  Translational  Sciences, Clinical and Translational Sciences  Award. SB, EA, TT and JG acknowledges Purdue Discovery Park Data Science Award \textit{"Fingerprints of the Human Brain: A Data Science Perspective"}. We also thank Dr. Kausar Abbas for useful comments and discussions.

The authors declare that they have no financial or non-financial conflicts of interest. All procedures performed in this study involving human participants were in accordance with the ethical standards of the Purdue Institutional Review Board and with the 1964 Helsinki declaration and its later amendments or comparable ethical standards. Informed consent was obtained from all participants prior to participation in the study.

\bibliographystyle{unsrt}
\bibliography{Mendeley.bib}

\begin{thebibliography}{10}

\bibitem{VanEssen2012ThePerspective}
D.~C. Van~Essen, K.~Ugurbil, E.~Auerbach, D.~Barch, T.~E.J. Behrens,
  R.~Bucholz, A.~Chang, L.~Chen, M.~Corbetta, S.~W. Curtiss, S.~Della~Penna,
  D.~Feinberg, M.~F. Glasser, N.~Harel, A.~C. Heath, L.~Larson-Prior,
  D.~Marcus, G.~Michalareas, S.~Moeller, R.~Oostenveld, S.~E. Petersen,
  F.~Prior, B.~L. Schlaggar, S.~M. Smith, A.~Z. Snyder, J.~Xu, and E.~Yacoub.
\newblock {The Human Connectome Project: A data acquisition perspective}.
\newblock {\em NeuroImage}, 62(4):2222--2231, 2012.

\bibitem{VanEssen2013TheOverview}
David~C. Van~Essen, Stephen~M. Smith, Deanna~M. Barch, Timothy~E.J. Behrens,
  Essa Yacoub, and Kamil Ugurbil.
\newblock {The WU-Minn Human Connectome Project: An overview}.
\newblock {\em NeuroImage}, 80:62--79, 2013.

\bibitem{Keator2016TheRepository}
David~B. Keator, Theo~G.M. van Erp, Jessica~A. Turner, Gary~H. Glover, Bryon~A.
  Mueller, Thomas~T. Liu, James~T. Voyvodic, Jerod Rasmussen, Vince~D. Calhoun,
  Hyo~Jong Lee, Arthur~W. Toga, Sarah McEwen, Judith~M. Ford, Daniel~H.
  Mathalon, Michele Diaz, Daniel~S. O'Leary, H.~Jeremy~Bockholt, Syam Gadde,
  Adrian Preda, Cynthia~G. Wible, Hal~S. Stern, Aysenil Belger, Gregory
  McCarthy, Burak Ozyurt, and Steven~G. Potkin.
\newblock {The Function Biomedical Informatics Research Network Data
  Repository}.
\newblock {\em NeuroImage}, 124:1074--1079, 2016.

\bibitem{Jack2008TheMethods}
Clifford~R. Jack, Matt~A. Bernstein, Nick~C. Fox, Paul Thompson, Gene
  Alexander, Danielle Harvey, Bret Borowski, Paula~J. Britson, Jennifer
  L.~Whitwell, Chadwick Ward, Anders~M. Dale, Joel~P. Felmlee, Jeffrey~L.
  Gunter, Derek~L.G. Hill, Ron Killiany, Norbert Schuff, Sabrina Fox-Bosetti,
  Chen Lin, Colin Studholme, Charles~S. DeCarli, Gunnar Gunnar~Krueger,
  Heidi~A. Ward, Gregory~J. Metzger, Katherine~T. Scott, Richard Mallozzi,
  Daniel Blezek, Joshua Levy, Josef~P. Debbins, Adam~S. Fleisher, Marilyn
  Albert, Robert Green, George Bartzokis, Gary Glover, John Mugler, and
  Michael~W. Weiner.
\newblock {The Alzheimer's disease neuroimaging initiative (ADNI): MRI
  methods}.
\newblock {\em Journal of Magnetic Resonance Imaging}, 27(4):685--691, 4 2008.

\bibitem{VanHorn2009MultisiteTrials.}
John~Darrell Van~Horn and Arthur~W Toga.
\newblock {Multisite neuroimaging trials.}
\newblock {\em Current opinion in neurology}, 22(4):370--8, 8 2009.

\bibitem{Friedman2006ReducingDifferences}
Lee Friedman, Gary~H. Glover, and {The FBIRN Consortium}.
\newblock {Reducing interscanner variability of activation in a multicenter
  fMRI study: Controlling for signal-to-fluctuation-noise-ratio (SFNR)
  differences}.
\newblock {\em NeuroImage}, 33(2):471--481, 11 2006.

\bibitem{Mulkern2008EstablishmentConsortium}
Robert~V. Mulkern, Peter Forbes, Kevin Dewey, Stravoula Osganian, Maureen
  Clark, Sharon Wong, Uma Ramamurthy, Larry Kun, and Tina~Young Poussaint.
\newblock {Establishment and Results of a Magnetic Resonance Quality Assurance
  Program for the Pediatric Brain Tumor Consortium}.
\newblock {\em Academic Radiology}, 15(9):1099--1110, 9 2008.

\bibitem{Brown2011MultisiteData}
Gregory~G. Brown, Daniel~H. Mathalon, Hal Stern, Judith Ford, Bryon Mueller,
  Douglas~N. Greve, Gregory McCarthy, James Voyvodic, Gary Glover, Michele
  Diaz, Elizabeth Yetter, I.~Burak Ozyurt, Kasper~W. Jorgensen, Cynthia~G.
  Wible, Jessica~A. Turner, Wesley~K. Thompson, and Steven~G. Potkin.
\newblock {Multisite reliability of cognitive BOLD data}.
\newblock {\em NeuroImage}, 54(3):2163--2175, 2011.

\bibitem{Voyvodic2006ActivationStrengths}
James~T. Voyvodic.
\newblock {Activation mapping as a percentage of local excitation: fMRI
  stability within scans, between scans and across field strengths}.
\newblock {\em Magnetic Resonance Imaging}, 24(9):1249--1261, 11 2006.

\bibitem{Friedman2008Test-retestStudy}
Lee Friedman, Hal Stern, Gregory~G. Brown, Daniel~H. Mathalon, Jessica Turner,
  Gary~H. Glover, Randy~L. Gollub, John Lauriello, Kelvin~O. Lim, Tyrone
  Cannon, Douglas~N. Greve, Henry~Jeremy Bockholt, Aysenil Belger, Bryon
  Mueller, Michael~J. Doty, Jianchun He, William Wells, Padhraic Smyth, Steve
  Pieper, Seyoung Kim, Marek Kubicki, Mark Vangel, and Steven~G. Potkin.
\newblock {Test-retest and between-site reliability in a multicenter fMRI
  study}.
\newblock {\em Human Brain Mapping}, 29(8):958--972, 2008.

\bibitem{Casey1998ReproductibilityTask}
B~J Casey, J~D Cohen, K~O'Craven, R~J Davidson, W~Irwin, C~A Nelson, D~C Noll,
  X~Hu, M~J Lowe, B~R Rosen, C~L Truwitt, and P~A Turski.
\newblock {Reproductibility of fMRI results across four institutions using a
  spatial working memory task}.
\newblock {\em Neuroimage}, 8(3):249--261, 1998.

\bibitem{Gountouna2010FunctionalTask}
Viktoria-Eleni Gountouna, Dominic~E. Job, Andrew~M. McIntosh, T.~William~J.
  Moorhead, G.~Katherine~L. Lymer, Heather~C. Whalley, Jeremy Hall, Gordon~D.
  Waiter, David Brennan, David~J. McGonigle, Trevor~S. Ahearn, Jonathan
  Cavanagh, Barrie Condon, Donald.~M. Hadley, Ian Marshall, Alison~D. Murray,
  J.~Douglas Steele, Joanna~M. Wardlaw, and Stephen~M. Lawrie.
\newblock {Functional Magnetic Resonance Imaging (fMRI) reproducibility and
  variance components across visits and scanning sites with a finger tapping
  task}.
\newblock {\em NeuroImage}, 49(1):552--560, 1 2010.

\bibitem{Suckling2008ComponentsPower}
John Suckling, David Ohlssen, Christopher Andrew, Glyn Johnson, Steven~C.R.
  Williams, Martin Graves, Chi-Hua Chen, David Spiegelhalter, and Ed~Bullmore.
\newblock {Components of variance in a multicentre functional MRI study and
  implications for calculation of statistical power}.
\newblock {\em Human Brain Mapping}, 29(10):1111--1122, 10 2008.

\bibitem{Yendiki2010Multi-siteIndices}
Anastasia Yendiki, Douglas~N. Greve, Stuart Wallace, Mark Vangel, Jeremy
  Bockholt, Bryon~A. Mueller, Vince Magnotta, Nancy Andreasen, Dara~S. Manoach,
  and Randy~L. Gollub.
\newblock {Multi-site characterization of an fMRI working memory paradigm:
  Reliability of activation indices}.
\newblock {\em NeuroImage}, 53(1):119--131, 10 2010.

\bibitem{Zou2005ReproducibilityNetwork}
Kelly~H. Zou, Douglas~N. Greve, Meng Wang, Steven~D. Pieper, Simon~K. Warfield,
  Nathan~S. White, Sanjay Manandhar, Gregory~G. Brown, Mark~G. Vangel, Ron
  Kikinis, and William~M. Wells.
\newblock {Reproducibility of Functional MR Imaging: Preliminary Results of
  Prospective Multi-institutional Study Performed by Biomedical Informatics
  Research Network}.
\newblock {\em Radiology}, 237(3):781--789, 12 2005.

\bibitem{Noble2017MultisiteConnectivity}
Stephanie Noble, Dustin Scheinost, Emily~S. Finn, Xilin Shen, Xenophon
  Papademetris, Sarah~C. McEwen, Carrie~E. Bearden, Jean Addington, Bradley
  Goodyear, Kristin~S. Cadenhead, Heline Mirzakhanian, Barbara~A. Cornblatt,
  Doreen~M. Olvet, Daniel~H. Mathalon, Thomas~H. McGlashan, Diana~O. Perkins,
  Aysenil Belger, Larry~J. Seidman, Heidi Thermenos, Ming~T. Tsuang, Theo~G.M.
  van Erp, Elaine~F. Walker, Stephan Hamann, Scott~W. Woods, Tyrone~D. Cannon,
  and R.~Todd Constable.
\newblock {Multisite reliability of MR-based functional connectivity}.
\newblock {\em NeuroImage}, 146:959--970, 2 2017.

\bibitem{Deprez2018Multi-centerMeasures}
S.~Deprez, Michiel~B. de~Ruiter, S.~Bogaert, R.~Peeters, J.~Belderbos,
  D.~De~Ruysscher, S.~Schagen, S.~Sunaert, P.~Pullens, and E.~Achten.
\newblock {Multi-center reproducibility of structural, diffusion tensor, and
  resting state functional magnetic resonance imaging measures}.
\newblock {\em Neuroradiology}, 60(6):617--634, 6 2018.

\bibitem{Noble2017InfluencesUtility}
Stephanie Noble, Marisa~N Spann, Fuyuze Tokoglu, Xilin Shen, R~Todd Constable,
  and Dustin Scheinost.
\newblock {Influences on the Test–Retest Reliability of Functional
  Connectivity MRI and its Relationship with Behavioral Utility}.
\newblock {\em Cerebral Cortex}, 27(11):5415--5429, 11 2017.

\bibitem{Jovicich2016LongitudinalStudy}
Jorge Jovicich, Ludovico Minati, Moira Marizzoni, Rocco Marchitelli, Roser
  Sala-Llonch, David Bartr{\'{e}}s-Faz, Jennifer Arnold, Jens Benninghoff, Ute
  Fiedler, Luca Roccatagliata, Agnese Picco, Flavio Nobili, Oliver Blin,
  Stephanie Bombois, Renaud Lopes, Régis Bordet, Julien Sein, Jean-Philippe
  Ranjeva, Mira Didic, Hélène Gros-Dagnac, Pierre Payoux, Giada Zoccatelli,
  Franco Alessandrini, Alberto Beltramello, Núria Bargall{\'{o}}, Antonio
  Ferretti, Massimo Caulo, Marco Aiello, Carlo Cavaliere, Andrea Soricelli,
  Lucilla Parnetti, Roberto Tarducci, Piero Floridi, Magda Tsolaki, Manos
  Constantinidis, Antonios Drevelegas, Paolo~Maria Rossini, Camillo Marra,
  Peter Sch{\"{o}}nknecht, Tilman Hensch, Karl-Titus Hoffmann, Joost~P. Kuijer,
  Pieter~Jelle Visser, Frederik Barkhof, and Giovanni~B. Frisoni.
\newblock {Longitudinal reproducibility of default-mode network connectivity in
  healthy elderly participants: A multicentric resting-state fMRI study}.
\newblock {\em NeuroImage}, 124:442--454, 1 2016.

\bibitem{Marchitelli2016Test-retestTechniques}
Rocco Marchitelli, Ludovico Minati, Moira Marizzoni, Beatriz Bosch, David
  Bartr{\'{e}}s-Faz, Bernhard~W. M{\"{u}}ller, Jens Wiltfang, Ute Fiedler, Luca
  Roccatagliata, Agnese Picco, Flavio Nobili, Oliver Blin, Stephanie Bombois,
  Renaud Lopes, Régis Bordet, Julien Sein, Jean-Philippe Ranjeva, Mira Didic,
  Hélène Gros-Dagnac, Pierre Payoux, Giada Zoccatelli, Franco Alessandrini,
  Alberto Beltramello, Núria Bargall{\'{o}}, Antonio Ferretti, Massimo Caulo,
  Marco Aiello, Carlo Cavaliere, Andrea Soricelli, Lucilla Parnetti, Roberto
  Tarducci, Piero Floridi, Magda Tsolaki, Manos Constantinidis, Antonios
  Drevelegas, Paolo~Maria Rossini, Camillo Marra, Peter Sch{\"{o}}nknecht,
  Tilman Hensch, Karl-Titus Hoffmann, Joost~P. Kuijer, Pieter~Jelle Visser,
  Frederik Barkhof, Giovanni~B. Frisoni, and Jorge Jovicich.
\newblock {Test-retest reliability of the default mode network in a
  multi-centric fMRI study of healthy elderly: Effects of data-driven
  physiological noise correction techniques}.
\newblock {\em Human Brain Mapping}, 37(6):2114--2132, 6 2016.

\bibitem{Huang2012ReproducibilityScanners}
Lejian Huang, Xue Wang, Marwan~N. Baliki, Lei Wang, A.~Vania Apkarian, and
  Todd~B. Parrish.
\newblock {Reproducibility of Structural, Resting-State BOLD and DTI Data
  between Identical Scanners}.
\newblock {\em PLoS ONE}, 7(10):e47684, 10 2012.

\bibitem{Turner2013ASchizophrenia}
Jessica~A Turner, Eswar Damaraju, Theo~G.M. Van~Erp, Daniel~H Mathalon,
  Judith~M Ford, James Voyvodic, Bryon~A. Mueller, Aysenil Belger, Juan
  Bustillo, Sarah~Christine McEwen, Steven~G. Potkin, Functional~Imaging BIRN,
  and Vince~D Calhoun.
\newblock {A multi-site resting state fMRI study on the amplitude of low
  frequency fluctuations in schizophrenia}.
\newblock {\em Frontiers in Neuroscience}, 7:137, 8 2013.

\bibitem{Feis2015ICA-basedFMRI}
Rogier~A. Feis, Stephen~M. Smith, Nicola Filippini, Gwenaëlle Douaud, Elise
  G.~P. Dopper, Verena Heise, Aaron~J. Trachtenberg, John~C. van Swieten,
  Mark~A. van Buchem, Serge A. R.~B. Rombouts, and Clare~E. Mackay.
\newblock {ICA-based artifact removal diminishes scan site differences in
  multi-center resting-state fMRI}.
\newblock {\em Frontiers in Neuroscience}, 9:395, 10 2015.

\bibitem{Jann2015FunctionalNetworks}
Kay Jann, Dylan~G. Gee, Emily Kilroy, Simon Schwab, Robert~X. Smith, Tyrone~D.
  Cannon, and Danny~J.J. Wang.
\newblock {Functional connectivity in BOLD and CBF data: Similarity and
  reliability of resting brain networks}.
\newblock {\em NeuroImage}, 106:111--122, 2 2015.

\bibitem{Biswal1995FunctionalMri}
Bharat Biswal, F.~Zerrin~Yetkin, Victor~M. Haughton, and James~S. Hyde.
\newblock {Functional connectivity in the motor cortex of resting human brain
  using echo-planar mri}.
\newblock {\em Magnetic Resonance in Medicine}, 34(4):537--541, 10 1995.

\bibitem{Fox2007SpontaneousImaging}
Michael~D. Fox and Marcus~E. Raichle.
\newblock {Spontaneous fluctuations in brain activity observed with functional
  magnetic resonance imaging}.
\newblock {\em Nature Reviews Neuroscience}, 8(9):700--711, 9 2007.

\bibitem{Greicius2003FunctionalHypothesis.}
Michael~D Greicius, Ben Krasnow, Allan~L Reiss, and Vinod Menon.
\newblock {Functional connectivity in the resting brain: a network analysis of
  the default mode hypothesis.}
\newblock {\em Proceedings of the National Academy of Sciences of the United
  States of America}, 100(1):253--8, 1 2003.

\bibitem{Beckmann2005InvestigationsAnalysis.}
Christian~F Beckmann, Marilena DeLuca, Joseph~T Devlin, and Stephen~M Smith.
\newblock {Investigations into resting-state connectivity using independent
  component analysis.}
\newblock {\em Philosophical transactions of the Royal Society of London.
  Series B, Biological sciences}, 360(1457):1001--13, 5 2005.

\bibitem{Shehzad2009TheReliable}
Zarrar Shehzad, A.~M.~Clare Kelly, Philip~T. Reiss, Dylan~G. Gee, Kristin
  Gotimer, Lucina~Q. Uddin, Sang~Han Lee, Daniel~S. Margulies, Amy~Krain Roy,
  Bharat~B. Biswal, Eva Petkova, F.~Xavier Castellanos, and Michael~P. Milham.
\newblock {The Resting Brain: Unconstrained yet Reliable}.
\newblock {\em Cerebral Cortex}, 19(10):2209--2229, 10 2009.

\bibitem{Mayer2011FunctionalInjury}
Andrew~R. Mayer, Maggie~V. Mannell, Josef Ling, Charles Gasparovic, and
  Ronald~A. Yeo.
\newblock {Functional connectivity in mild traumatic brain injury}.
\newblock {\em Human Brain Mapping}, 32(11):1825--1835, 11 2011.

\bibitem{Broyd2009Default-modeReview}
Samantha~J. Broyd, Charmaine Demanuele, Stefan Debener, Suzannah~K. Helps,
  Christopher~J. James, and Edmund~J.S. Sonuga-Barke.
\newblock {Default-mode brain dysfunction in mental disorders: A systematic
  review}.
\newblock {\em Neuroscience {\&} Biobehavioral Reviews}, 33(3):279--296, 3
  2009.

\bibitem{Contreras2015TheAdults}
Joey~A. Contreras, Joaquín Go{\~{n}}i, Shannon~L. Risacher, Olaf Sporns, and
  Andrew~J. Saykin.
\newblock {The Structural and Functional Connectome and Prediction of Risk for
  Cognitive Impairment in Older Adults}.
\newblock {\em Current Behavioral Neuroscience Reports}, 2(4):234--245, 12
  2015.

\bibitem{Bullmore2009ComplexSystems}
Ed~Bullmore and Olaf Sporns.
\newblock {Complex brain networks: graph theoretical analysis of structural and
  functional systems}.
\newblock {\em Nature Reviews Neuroscience}, 10(3):186--198, 3 2009.

\bibitem{Rubinov2010ComplexInterpretations}
Mikail Rubinov and Olaf Sporns.
\newblock {Complex network measures of brain connectivity: Uses and
  interpretations}.
\newblock {\em NeuroImage}, 52(3):1059--1069, 2010.

\bibitem{Sporns2014ContributionsNeuroscience}
Olaf Sporns.
\newblock {Contributions and challenges for network models in cognitive
  neuroscience}.
\newblock {\em Nature Neuroscience}, 17(5):652--660, 5 2014.

\bibitem{He2009NeuronalDisease}
Yong He, Zhang Chen, Gaolang Gong, and Alan Evans.
\newblock {Neuronal networks in Alzheimer's disease}, 8 2009.

\bibitem{Fornito2016FundamentalsAnalysis}
Alex Fornito, Andrew Zalesky, and Edward~T. Bullmore.
\newblock {\em {Fundamentals of brain network analysis}}.
\newblock Academic Press, 2016.

\bibitem{Braun2012TestretestMeasures}
Urs Braun, Michael~M. Plichta, Christine Esslinger, Carina Sauer, Leila Haddad,
  Oliver Grimm, Daniela Mier, Sebastian Mohnke, Andreas Heinz, Susanne Erk,
  Henrik Walter, Nina Seiferth, Peter Kirsch, and Andreas Meyer-Lindenberg.
\newblock {Test–retest reliability of resting-state connectivity network
  characteristics using fMRI and graph theoretical measures}.
\newblock {\em NeuroImage}, 59(2):1404--1412, 1 2012.

\bibitem{Wang2011GraphData}
Jin-Hui Wang, Xi-Nian Zuo, Suril Gohel, Michael~P. Milham, Bharat~B. Biswal,
  and Yong He.
\newblock {Graph Theoretical Analysis of Functional Brain Networks: Test-Retest
  Evaluation on Short- and Long-Term Resting-State Functional MRI Data}.
\newblock {\em PLoS ONE}, 6(7):e21976, 7 2011.

\bibitem{Schultz2016HigherReconfiguration}
Douglas~H. Schultz and Michael~W. Cole.
\newblock {Higher Intelligence Is Associated with Less Task-Related Brain
  Network Reconfiguration}.
\newblock {\em The Journal of Neuroscience}, 36(33):8551--8561, 2016.

\bibitem{Mira-Dominguez2014Connectotyping:Connectome}
Oscar Mira-Dominguez, Brian~D. Mills, Samuel~D. Carpenter, Kathleen~A. Grant,
  Christopher~D. Kroenke, Joel~T. Nigg, and Damien~A. Fair.
\newblock {Connectotyping: Model based fingerprinting of the functional
  connectome}.
\newblock {\em PLoS ONE}, 9(11), 2014.

\bibitem{Finn2015FunctionalConnectivity}
Emily~S. Finn, Xilin Shen, Dustin Scheinost, Monica~D. Rosenberg, Jessica
  Huang, Marvin~M. Chun, Xenophon Papademetris, and R.~Todd Constable.
\newblock {Functional connectome fingerprinting: Identifying individuals using
  patterns of brain connectivity}.
\newblock {\em Nature Neuroscience}, 18(11):1664--1671, 2015.

\bibitem{Finn2017CanConnectivity}
Emily~S. Finn, Dustin Scheinost, Daniel~M. Finn, Xilin Shen, Xenophon
  Papademetris, and R.~Todd Constable.
\newblock {Can brain state be manipulated to emphasize individual differences
  in functional connectivity?}
\newblock {\em NeuroImage}, 160(March):140--151, 2017.

\bibitem{Gordon2017Individual-specificCorrelations}
Evan~M. Gordon, Timothy~O. Laumann, Babatunde Adeyemo, Adrian~W. Gilmore,
  Steven~M. Nelson, Nico~U.F. Dosenbach, and Steven~E. Petersen.
\newblock {Individual-specific features of brain systems identified with
  resting state functional correlations}.
\newblock {\em NeuroImage}, 146:918--939, 2017.

\bibitem{Waller2017EvaluatingFingerprints}
Lea Waller, Henrik Walter, Johann~D. Kruschwitz, Lucia Reuter, Sabine
  M{\"{u}}ller, Susanne Erk, and Ilya~M. Veer.
\newblock {Evaluating the replicability, specificity, and generalizability of
  connectome fingerprints}.
\newblock {\em NeuroImage}, 158:371--377, 9 2017.

\bibitem{Gordon2017PrecisionBrains}
Evan~M. Gordon, Timothy~O. Laumann, Adrian~W. Gilmore, Dillan~J. Newbold,
  Deanna~J. Greene, Jeffrey~J. Berg, Mario Ortega, Catherine Hoyt-Drazen,
  Caterina Gratton, Haoxin Sun, Jacqueline~M. Hampton, Rebecca~S. Coalson,
  Annie~L. Nguyen, Kathleen~B. McDermott, Joshua~S. Shimony, Abraham~Z. Snyder,
  Bradley~L. Schlaggar, Steven~E. Petersen, Steven~M. Nelson, and Nico~U.F.
  Dosenbach.
\newblock {Precision Functional Mapping of Individual Human Brains}.
\newblock {\em Neuron}, 95(4):791--807, 2017.

\bibitem{Zimmermann2018Subject-SpecificityConnectivity}
J.~Zimmermann, J.~Griffiths, M.~Schirner, P.~Ritter, and A.R. McIntosh.
\newblock {Subject-Specificity of the Correlation Between Large-Scale
  Structural and Functional Connectivity}.
\newblock {\em Network Neuroscience}, pages 1--35, 4 2018.

\bibitem{Satterthwaite2018PersonalizedNetworks}
Theodore~D. Satterthwaite, Cedric~H. Xia, and Danielle~S. Bassett.
\newblock {Personalized Neuroscience: Common and Individual-Specific Features
  in Functional Brain Networks}.
\newblock {\em Neuron}, 98(2):243--245, 4 2018.

\bibitem{Greene2018Task-inducedTraits}
Abigail~S. Greene, Siyuan Gao, Dustin Scheinost, and R.~Todd Constable.
\newblock {Task-induced brain state manipulation improves prediction of
  individual traits}.
\newblock {\em Nature Communications}, 9(1):2807, 12 2018.

\bibitem{Gonzalez-Castillo2017Task-basedQuestions}
Javier Gonzalez-Castillo and Peter~A. Bandettini.
\newblock {Task-based dynamic functional connectivity: Recent findings and open
  questions}.
\newblock {\em NeuroImage}, (August):1--8, 2017.

\bibitem{Amico2018TheConnectomes}
Enrico Amico and Joaquín Go{\~{n}}i.
\newblock {The quest for identifiability in human functional connectomes}.
\newblock {\em Scientific Reports}, 8(1):8254, 12 2018.

\bibitem{Pallares2018ExtractingConnectivity}
Vicente Pallar{\'{e}}s, Andrea Insabato, Ana Sanju{\'{a}}n, Simone K{\"{u}}hn,
  Dante Mantini, Gustavo Deco, and Matthieu Gilson.
\newblock {Extracting orthogonal subject- and condition-specific signatures
  from fMRI data using whole-brain effective connectivity}.
\newblock {\em NeuroImage}, 178:238--254, 9 2018.

\bibitem{Svaldi2018TowardsConnectomes}
Diana~O. Svaldi, Joaquín Go{\~{n}}i, Apoorva Bharthur~Sanjay, Enrico Amico,
  Shannon~L. Risacher, John~D. West, Mario Dzemidzic, Andrew Saykin, and Liana
  Apostolova.
\newblock {Towards Subject and Diagnostic Identifiability in the Alzheimer’s
  Disease Spectrum Based on Functional Connectomes}.
\newblock pages 74--82. Springer, Cham, 9 2018.

\bibitem{Vanderwal2017IndividualConditions}
Tamara Vanderwal, Jeffrey Eilbott, Emily~S. Finn, R.~Cameron Craddock, Adam
  Turnbull, and F.~Xavier Castellanos.
\newblock {Individual differences in functional connectivity during
  naturalistic viewing conditions}.
\newblock {\em NeuroImage}, 157:521--530, 8 2017.

\bibitem{Shen2017UsingConnectivity}
Xilin Shen, Emily~S Finn, Dustin Scheinost, Monica~D Rosenberg, Marvin~M Chun,
  Xenophon Papademetris, and R~Todd Constable.
\newblock {Using connectome-based predictive modeling to predict individual
  behavior from brain connectivity}.
\newblock {\em Nature Protocols}, 12(3):506--518, 2 2017.

\bibitem{Yoo2018Connectome-basedDatasets}
Kwangsun Yoo, Monica~D. Rosenberg, Wei-Ting Hsu, Sheng Zhang, Chiang-Shan~R.
  Li, Dustin Scheinost, R.~Todd Constable, and Marvin~M. Chun.
\newblock {Connectome-based predictive modeling of attention: Comparing
  different functional connectivity features and prediction methods across
  datasets}.
\newblock {\em NeuroImage}, 167:11--22, 2 2018.

\bibitem{Shah2016ReliabilityState}
Lubdha~M. Shah, Justin~A. Cramer, Michael~A. Ferguson, Rasmus~M. Birn, and
  Jeffrey~S. Anderson.
\newblock {Reliability and reproducibility of individual differences in
  functional connectivity acquired during task and resting state}.
\newblock {\em Brain and Behavior}, 6(5), 5 2016.

\bibitem{Cox1996AFNI:Neuroimages}
Robert~W Cox.
\newblock {AFNI: Software for Analysis and Visualization of Functional Magnetic
  Resonance Neuroimages}.
\newblock {\em COMPUTERS AND BIOMEDICAL RESEARCH}, 29:162--173, 1996.

\bibitem{Jenkinson2012FSL}
Mark Jenkinson, Christian~F Beckmann, Timothy E~J Behrens, Mark~W Woolrich, and
  Stephen~M Smith.
\newblock {FSL}.
\newblock 2012.

\bibitem{Smith2004AdvancesFSL}
Stephen~M Smith, Mark Jenkinson, Mark~W Woolrich, Christian~F Beckmann,
  Timothy~E.J. Behrens, Heidi Johansen-Berg, Peter~R Bannister, Marilena
  De~Luca, Ivana Drobnjak, David~E Flitney, Rami~K Niazy, James Saunders, John
  Vickers, Yongyue Zhang, Nicola De~Stefano, J~Michael Brady, and Paul~M
  Matthews.
\newblock {Advances in functional and structural MR image analysis and
  implementation as FSL}.
\newblock In {\em NeuroImage}, volume~23, 2004.

\bibitem{Amico2017MappingConsciousness}
Enrico Amico, Daniele Marinazzo, Carol Di~Perri, Lizette Heine, Jitka Annen,
  Charlotte Martial, Mario Dzemidzic, Murielle Kirsch, Vincent Bonhomme, Steven
  Laureys, and Joaquín Go{\~{n}}i.
\newblock {Mapping the functional connectome traits of levels of
  consciousness}.
\newblock {\em NeuroImage}, 148:201--211, 3 2017.

\bibitem{Coupe2010RobustImages.}
Pierrick Coup{\'{e}}, José~V Manj{\'{o}}n, Elias Gedamu, Douglas Arnold,
  Montserrat Robles, and D~Louis Collins.
\newblock {Robust Rician noise estimation for MR images.}
\newblock {\em Medical image analysis}, 14(4):483--93, 8 2010.

\bibitem{Coupe2008AnImages}
P.~Coupe, P.~Yger, S.~Prima, P.~Hellier, C.~Kervrann, and C.~Barillot.
\newblock {An Optimized Blockwise Nonlocal Means Denoising Filter for 3-D
  Magnetic Resonance Images}.
\newblock {\em IEEE Transactions on Medical Imaging}, 27(4):425--441, 4 2008.

\bibitem{Wiest-Daessle2008RicianDT-MRI}
Nicolas Wiest-Daessl{\'{e}}, Sylvain Prima, Pierrick Coup{\'{e}}, Sean~Patrick
  Morrissey, and Christian Barillot.
\newblock {Rician Noise Removal by Non-Local Means Filtering for Low
  Signal-to-Noise Ratio MRI: Applications to DT-MRI}.
\newblock pages 171--179. Springer, Berlin, Heidelberg, 2008.

\bibitem{Jones2010SourcesDisorder}
Tyler~B. Jones, Peter~A. Bandettini, Lauren Kenworthy, Laura~K. Case, Shawn~C.
  Milleville, Alex Martin, and Rasmus~M. Birn.
\newblock {Sources of group differences in functional connectivity: An
  investigation applied to autism spectrum disorder}.
\newblock {\em NeuroImage}, 49(1):401--414, 1 2010.

\bibitem{Power2012SpuriousMotion}
Jonathan~D. Power, Kelly~A. Barnes, Abraham~Z. Snyder, Bradley~L. Schlaggar,
  and Steven~E. Petersen.
\newblock {Spurious but systematic correlations in functional connectivity MRI
  networks arise from subject motion}.
\newblock {\em NeuroImage}, 59(3):2142--2154, 2 2012.

\bibitem{Power2014MethodsFMRI}
Jonathan~D. Power, Anish Mitra, Timothy~O. Laumann, Abraham~Z. Snyder,
  Bradley~L. Schlaggar, and Steven~E. Petersen.
\newblock {Methods to detect, characterize, and remove motion artifact in
  resting state fMRI}.
\newblock {\em NeuroImage}, 84:320--341, 1 2014.

\bibitem{Jo2010MappingRemoval.}
Hang~Joon Jo, Ziad~S Saad, W~Kyle Simmons, Lydia~A Milbury, and Robert~W Cox.
\newblock {Mapping sources of correlation in resting state FMRI, with artifact
  detection and removal.}
\newblock {\em NeuroImage}, 52(2):571--82, 8 2010.

\bibitem{Shen2013GroupwiseIdentification}
X~Shen, F~Tokoglu, X~Papademetris, and R~T Constable.
\newblock {Groupwise whole-brain parcellation from resting-state fMRI data for
  network node identification}.
\newblock {\em NeuroImage}, 82:403--415, 2013.

\bibitem{ThomasYeo2011TheConnectivity}
B.~T. Thomas~Yeo, Fenna~M. Krienen, Jorge Sepulcre, Mert~R. Sabuncu, Danial
  Lashkari, Marisa Hollinshead, Joshua~L. Roffman, Jordan~W. Smoller, Lilla
  Z{\"{o}}llei, Jonathan~R. Polimeni, Bruce Fischl, Hesheng Liu, and Randy~L.
  Buckner.
\newblock {The organization of the human cerebral cortex estimated by intrinsic
  functional connectivity}.
\newblock {\em Journal of Neurophysiology}, 106(3):1125--1165, 9 2011.

\bibitem{Team2013R:Computing}
R~core Team.
\newblock {R: A Language and Environment for Statistical Computing}, 2013.

\bibitem{Shrout1979IntraclassReliability.}
Patrick~E. Shrout and Joseph~L. Fleiss.
\newblock {Intraclass correlations: Uses in assessing rater reliability.}
\newblock {\em Psychological Bulletin}, 86(2):420--428, 1979.

\bibitem{McGraw1996FormingCoefficients.}
Kenneth~O. McGraw and S.~P. Wong.
\newblock {Forming inferences about some intraclass correlation coefficients.}
\newblock {\em Psychological Methods}, 1(1):30--46, 1996.

\bibitem{Yeo2011TheConnectivity}
B.T.~Thomas Yeo, Fenna~M. Krienen, Jorge Sepulcre, Mert~R. Sabuncu,
  D.~Lashkari, Marisa Hollinshead, Joshua~L. Roffman, Jordan~W. Smoller, Lilla
  Zollei, Jonathan~R. Polimeni, Bruce Fischl, Hesheng Liu, and Randy~L.
  Buckner.
\newblock {The organization of the human cerebral cortex estimated by intrinsic
  functional connectivity}.
\newblock {\em Journal of neurophysiology}, 106:1125--1165, 2011.

\bibitem{Mars2018ConnectivitySpaces}
Rogier~B. Mars, Richard~E. Passingham, and Saad Jbabdi.
\newblock {Connectivity Fingerprints: From Areal Descriptions to Abstract
  Spaces}.
\newblock {\em Trends in Cognitive Sciences}, 22(11):1026--1037, 11 2018.

\bibitem{Sripada2019BasicConnectomes}
Chandra Sripada, Mike Angstadt, Saige Rutherford, Daniel Kessler, Yura Kim,
  Mike Yee, and Elizaveta Levina.
\newblock {Basic Units of Inter-Individual Variation in Resting State
  Connectomes}.
\newblock {\em Scientific Reports}, 9(1):1900, 12 2019.

\end{thebibliography}
\newpage
\section*{Supplementary Material}
\beginsupplement


\begin{figure}[h!]
\centering
\includegraphics[scale =0.8,trim= {3cm 8cm 2cm 9cm}, clip=true]{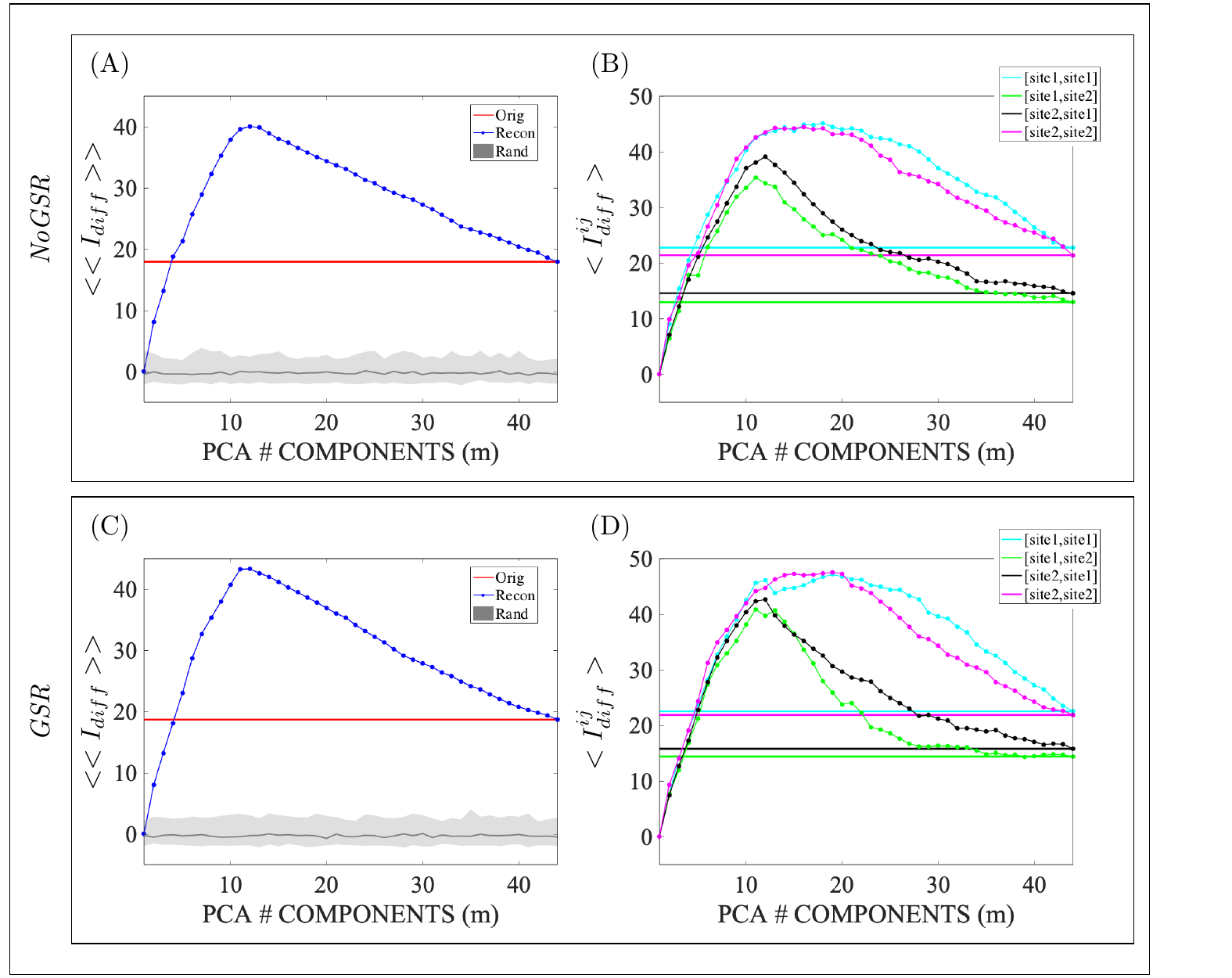}
\caption{Yale dataset. Multi-site differential identifiability $(<<I_{diff}>>*100)$ and differential identifiability of each [site\textit{i}, site\textit{j}] pair, $(<I_{diff}^{ij}>*100)$ for training data as a function of the number of principal components (PCs) used for reconstruction for resting-state data without global signal regression (\textit{NoGSR}; (A) and (B)); and with global signal regression (\textit{GSR}; (C) and (D)). In all figures solid lines denote $<<I_{diff}>>$ and $<I_{diff}^{ij}>$ as computed from the original FCs, whereas lines with circles denote the differential identifiability for reconstructed FCs as a function of \textbf{m}, the included number of components. In (A) and (C), the gray (shaded) area denotes the 95\% confidence interval for $<<I_{diff}>>$ over 100 random permutations of the test-retest FC pairs at each value of \textbf{m}. 
It may be observed that the benefit of reconstruction on differential identifiability was not dependent on the exclusion/inclusion of global signal regression.}
\label{Idiff_yale}
\end{figure}

\begin{table}[h!]
\centering
\caption{Yale dataset. Maximum percentage differential identifiability ($<I_{diff}^{ij*}>*100$), explained variance ($R^{2}$) and number of principal components for each [site\textit{i}, site\textit{j}] pair ($m^{ij*}$) for Training datasets without global signal regression (\textit{NoGSR}) and with global signal regression (\textit{GSR}).} 
\begin{tabular}{|c|c|c|c|c|}
\hline
                       & [site\textit{i}, site\textit{j}] & $<I_{diff}^{ij*}>$ & $m^{ij*}$  & $R^{2}$ \\ \hline
\multirow{4}{*}{\textit{NoGSR}} & [site1, site1]        & 45.1     & 18  & 0.75\\ \cline{2-5} 
                       & [site1, site2]        & 35.3     & 11  &  0.63\\ \cline{2-5} 
                       & [site2, site1]        & 39.1     & 12  & 0.65\\ \cline{2-5} 
                       & [site2, site2]        & 44.4     & 16 & 0.72\\ \hline
\multirow{4}{*}{\textit{GSR}}   & [site1, site1]        & 47.1     & 19  & 0.76\\ \cline{2-5} 
                       & [site1, site2]        & 40.9     & 11  & 0.64\\ \cline{2-5} 
                       & [site2, site1]        & 42.6     & 12  & 0.66\\ \cline{2-5} 
                       & [site2, site2]        & 47.5     & 19  & 0.76\\ \hline
\end{tabular}
\label{I_diff_ij_yale}
\end{table}

\begin{figure}[h!]
\centering
\includegraphics[scale =0.8,trim= {2.8cm 9cm 2cm 10cm}, clip=true]{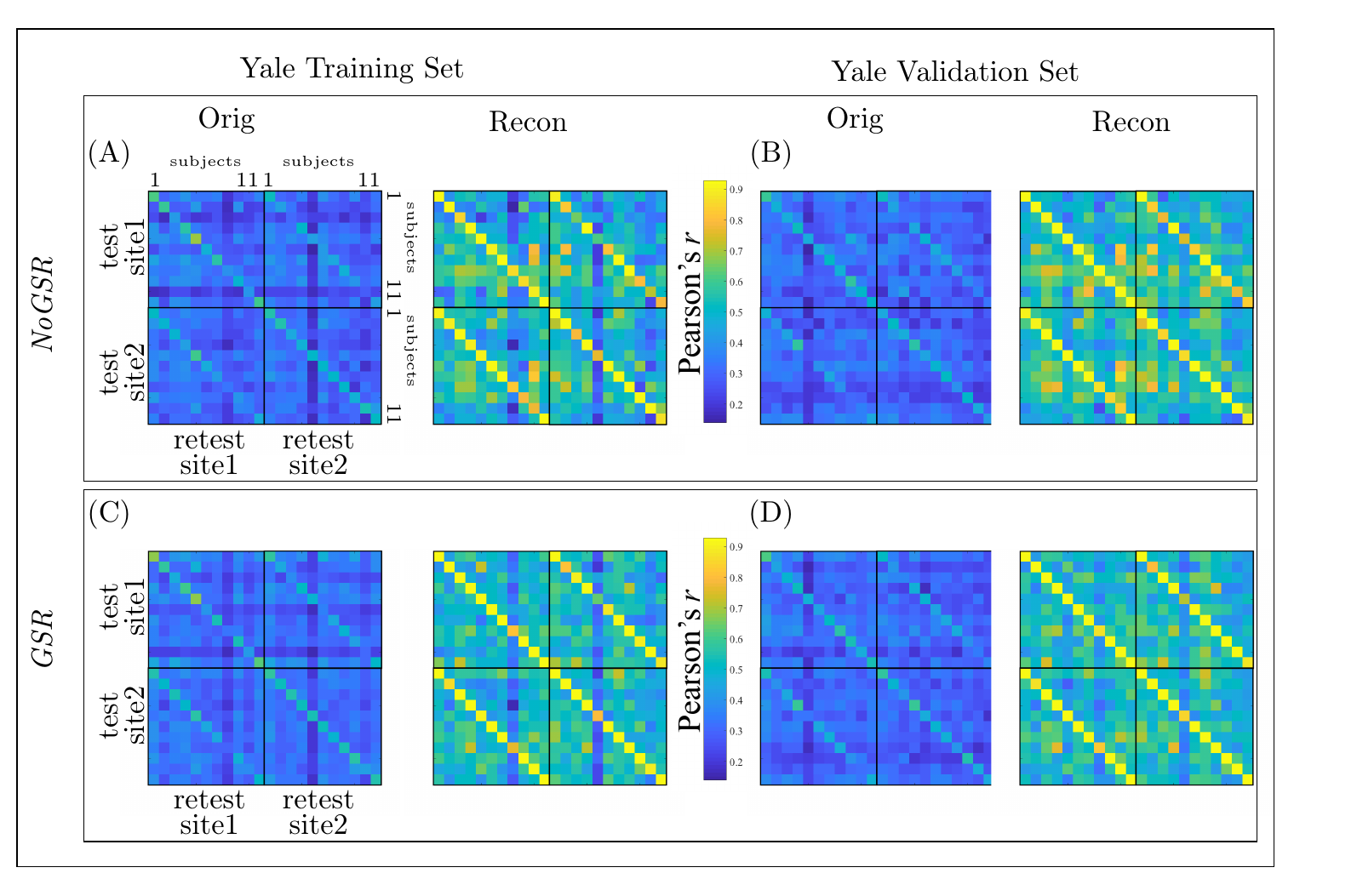}
\caption{Yale dataset. Identifiability matrices (\textbf{I}) of the original (Orig) and reconstructed (Recon) data for the \textit{Training}, (A) and (C), and \textit{Validation}, (B) and (D) sets of resting-state functional connectomes without global signal regression (\textit{NoGSR}; (A) and (B)) and with global signal regression (\textit{GSR}; (C) and (D)). The Identifiability matrix (\textbf{I}) has a blockwise structure where each block is $I^{ij}$, representing the identifiability for the [site\textit{i}, site\textit{j}] pair.
Note that identifiability was meaningfully improved across sites regardless of the exclusion/inclusion of global signal regression.}
\label{Ident_matrices_yale}
\end{figure}

\end{document}